\documentclass[12pt]{article}
\usepackage[dvipdfm]{graphicx}
\usepackage{bm}
\usepackage{amsmath}
\usepackage{amssymb}
\usepackage{float}
\usepackage{url}

\renewcommand{\thepage}{}
\makeatletter
\@addtoreset{equation}{section}
\renewcommand{\theequation}{\thesection.\@arabic\c@equation}
\makeatother
\renewcommand{\thefootnote}{\fnsymbol{footnote}}
\begin{document}
\begin{titlepage}
\title{
\vspace*{-4ex}
\hfill
\begin{minipage}{3.5cm}
\end{minipage}\\
\bf Numerical Solutions of Open String Field Theory
in Marginally Deformed Backgrounds
\vspace{0.5em}
}

\author{Isao {\sc Kishimoto}$^{1}$
\ \ and\ \  
Tomohiko {\sc Takahashi}$^{2}$
\\
\vspace{0.5ex}\\
$^1${\it Faculty of Education, Niigata University,}\\
{\it Niigata 950-2181, Japan}\\
$^2${\it Department of Physics, Nara Women's University,}\\
{\it Nara 630-8506, Japan}}
\date{June, 2013}
\maketitle
%

\begin{abstract}
\normalsize

We investigate numerical solutions of bosonic open string field theory
in some marginally deformed backgrounds, which are obtained by
expanding the action around an identity-based marginal solution with one
parameter. We construct numerical solutions in the Siegel gauge and the
 Landau gauge corresponding to the tachyon vacuum.
Their vacuum energy approximately cancels the D-brane tension
for larger intervals of the parameter with increasing truncation level. 
The result is consistent with the previous expectation that the
identity-based marginal solution has vanishing energy 
regardless of the values of the parameter.
We also study the marginal branch ($M$-branch) and the vacuum branch
($V$-branch) and evaluate not only the vacuum energy but also the gauge
invariant overlaps with the graviton and the closed tachyon.
We observe that there is a finite bound for the value of the massless field 
of numerical solutions even in the marginally deformed background.

\end{abstract}
\end{titlepage}

\renewcommand{\thepage}{\arabic{page}}
\renewcommand{\thefootnote}{\arabic{footnote}}
\setcounter{page}{1}
\setcounter{footnote}{0}
%
\section{Introduction}

Open bosonic string field theory has a non-perturbative vacuum
corresponding to a marginal deformation such as background Wilson lines
\cite{Sen:2000hx}. 
The effective potential of the massless field becomes increasingly 
flat as the truncation level is increased. 
On the analytical side, there are classical solutions expected
to represent marginal deformations.
As such solutions, identity-based marginal solutions were constructed in 
\cite{Takahashi:2001pp,Takahashi:2002ez,Kishimoto:2005bs}.
Furthermore, other types of marginal solutions have been constructed
\cite{Schnabl:2007az,Kiermaier:2007ba,Fuchs:2007yy,Kiermaier:2007vu}.
The vacuum energies of these solutions are formally proved to be zero
by differentiating and integrating the action with respect to a 
deformation parameter.
However, it may be provided as a sort of indefinite
quantity, especially in the case of identity-based solutions.

The tachyon vacuum exists even in the presence of Wilson lines, and
the vacuum energy is expected to cancel the D-brane tension, which
is equivalent to that of no Wilson lines.
If we expand the string field around an analytic solution
corresponding to background Wilson lines, 
the action for the fluctuation describes strings on the Wilson line
background.
Accordingly, the expanded theory should have a non-perturbative vacuum,
the vacuum energy of which is given as the same one without Wilson
lines.
Actually, analytic tachyon vacuum solutions in the theory expanded around 
identity-based marginal solutions have been constructed in \cite{Inatomi:2012nv}
using the ``$K'Bc$ algebra'' and it is shown that their vacuum energy cancels a D-brane tension.
This provides evidence that the vacuum energy of the identity-based 
marginal solutions is zero.

Here, we construct numerical tachyon vacuum 
solutions, which satisfy other gauge conditions:
the Siegel gauge and the Landau gauge, using the level truncation method
in the theory around an
identity-based marginal solution with one real parameter $x$.
We find that their vacuum energy approximately cancels
the value of a D-brane tension for larger intervals of the parameter with
increasing level.\footnote{
Precisely, in the Landau gauge, the numerical behavior may not be stable for large $|x|$.
}
The result is consistent with that of the analytic approach in \cite{Inatomi:2012nv}
and implies that the energy of the identity-based marginal solution vanishes.

In \cite{Sen:2000hx}, it was observed that there are two branches
for an effective potential of a constant mode of  the massless field, 
denoted as $a_s$, with the level truncation approximation.
One is a ``marginal branch'' ($M$-branch), 
which includes the trivial zero solution,
and the other is a ``vacuum branch'' ($V$-branch), which includes
the tachyon vacuum solution.
With increasing level, the shape of the $M$-branch becomes flatter.
However, at a finite value of $a_s$, the $M$-branch and the $V$-branch 
merge and there is a maximum value of $a_s$ for the $M$-branch.
Recently, such a phenomenon was also observed 
for further higher level computations in \cite{Kudrna:2012um}.
In this context, we investigate the $M$-branch and the $V$-branch
in the theory around the identity-based marginal solution.
We find that the graph of the potential moves to the horizontal direction
for small values of the parameter $|x|$.
As for the $V$-branch, we observe that the value of $|a_s|$ at the
potential minimum has a finite bound around 0.3.
On the other hand, the $M$-branch seems to be unstable for large values
of $|x|$.

We evaluate not only the vacuum energy as mentioned above 
but also gauge invariant overlaps
with the graviton and the closed tachyon for the numerical solutions obtained.
For the tachyon vacuum (the minimum of the $V$-branch) in the theory around 
the identity-based marginal solution, 
we find that, with increasing level, the gauge invariant overlap with
the graviton approaches $1$ for various values of $x$
but that with the closed tachyon approaches $e^{-4ix}$.
On the other hand, for the $M$-branch in the original theory,
the gauge invariant overlap with the graviton approaches $0$ for various
values of $a_s$ but that with the closed tachyon 
depends on $a_s$, 
such as $1-e^{-ica_s}$ with some constant $c$ approximately.

This paper is organized as follows.
In \S \ref{sec:E_x}, we will construct numerical solutions,
both in the Siegel and Landau gauges,
 in the theory expanded around 
an identity-based marginal solution with one parameter $x$,
and evaluate their gauge invariants.
In \S\ref{sec:MV}, we will discuss the $M$-branch and $V$-branch
in the expanded theory for various values of $x$.
In \S\ref{sec:GIO}, we will comment on the gauge invariant overlaps
with the graviton and the closed tachyon
for numerical solutions in the $M$-branch.
In \S\ref{sec:RM}, we will give some concluding remarks.
In appendix \ref{sec:BRST}, we will show some numerical results on 
the BRST invariance  of the solutions.

\section{Tachyon vacuum around an identity-based marginal solution
\label{sec:E_x}
}

The equation of motion in open bosonic string field theory is given by
$Q_{\rm B}\Psi+\Psi*\Psi=0$.
As an analytic classical solution, we have 
a type of identity-based solution \cite{Takahashi:2001pp,Takahashi:2002ez,Kishimoto:2005bs}:
\begin{align}
\Psi_0=-\int_{C_{\rm left}}\frac{dz}{2\pi i}\frac{i}{2\sqrt{\alpha'}}
F(z)c(z)\partial X^{25}(z)I+\frac{1}{4}\int_{C_{\rm left}}\frac{dz}{2\pi i}
F(z)^2c(z)I,
\label{Psi0id}
\end{align}
where $I$ is the identity string field and 
$F(z)$ is a function that satisfies $F(-1/z)=z^2F(z)$.
In the integrations, $C_{\rm left}$
denotes the path along a unit half circle such as ${\rm Re}\,z\ge 0$. 
We can see that this
solution corresponds to the Wilson line along the 25th direction from the
study of the 
expanded theory around the solution. In this solution, the Wilson line
parameter is involved as
\begin{align}
 f=\int_{C_{\rm left}}\frac{dz}{2\pi i} F(z).
\label{f=intF}
\end{align}
The usual Wilson line is proportional to this quantity. 
Other modes of the function can be gauged away \cite{Kishimoto:2005bs}.

Expanding the string field around the solution as $\Psi=\Psi_0+\Phi$,
we can find the action $S'[\Phi]$ for the fluctuation around the Wilson line
background:
\begin{align}
S'[\Phi]\equiv S[\Psi_0+\Phi]-S[\Psi_0]
= -\!\left(\frac{1}{2}\langle\Phi ,Q'\Phi\rangle\!+\!\frac{1}{3}\langle\Phi,\Phi*\Phi\rangle\right).
\label{S'Phi}
\end{align}
The modified BRST operator in the expanded action is given
by
\begin{align}
Q'=Q_{\rm B}-\oint\frac{dz}{2\pi i}\frac{i}{2\sqrt{\alpha'}}
F(z)c(z)\partial X^{25}(z)+\frac{1}{4}\oint\frac{dz}{2\pi i}
F(z)^2c(z),
\label{Q'}
\end{align}
where $Q_{\rm B}$ denotes the
original BRST operator and the integration contour is along the unit circle.
In the following, we take a function $F(z)$ as
$F(z)=-x(z+1/z)z^{-1}$
for simplicity, where $x$ is a real parameter.
 Then,  (\ref{Q'}) is explicitly written as
\begin{align}
Q'=Q_{\rm B}+
\frac{x}{\sqrt{2}}\sum_{n\in \mathbb{Z}} c_n(\alpha_{-n-1}^{25}+\alpha_{-n+1}^{25})+\frac{x^2}{4}(2c_0+c_{-2}+c_2).
\label{Q'x}
\end{align}
With respect to the above $Q'$, we solve the equation of motion:
\begin{align}
Q'\Phi+\Phi*\Phi=0,
\label{EOMQ'}
\end{align}
numerically.

First of all, 
we construct the tachyon vacuum solution in the Siegel and Landau gauges,
which corresponds to the analytic solution constructed by 
the method of $K'Bc$ algebra in \cite{Inatomi:2012nv}:
\begin{align}
\Phi_T&=\frac{1}{\sqrt{1+K'}}(c+cK'Bc)\frac{1}{\sqrt{1+K'}},
\label{PhiT}
\end{align}
which satisfies the other gauge condition.\footnote{
It satisfies a kind of 
``dressed ${\cal B}_0$ gauge'' condition \cite{Erler:2009uj}:
\begin{align}
\frac{1}{\sqrt{1+K'}}\left[({\cal B}_0-{\cal B}_0^{\dagger})
\left(
\sqrt{1+K'}\,\Phi_T\,\sqrt{1+K'}
\right)\right]\frac{1}{\sqrt{1+K'}}=0.
\end{align}
}
To construct a numerical solution to the equation of motion (\ref{EOMQ'})
with a gauge condition, 
we solve
\begin{align}
&{\cal P}_1\Phi=0,
\label{G-cond}\\
&{\cal P}_2(Q'\Phi+\Phi*\Phi)=0,
\label{P2EOM}
\end{align}
for $\Phi$. 
${\cal P}_1$ and ${\cal P}_2=1-{\rm bpz}({\cal P}_1)$ are
projections determined by a gauge condition.
In the case of the Siegel gauge, these are given by
\begin{align}
&{\cal P}_1={\cal P}_2=c_0b_0,
\end{align}
and, in the case of the Landau gauge \cite{Asano:2006hk, Asano:2008iu},
these are\footnote{
For the ghost number $1$ string fields, the condition ${\cal P}_1\Phi=0$
can be rewritten as
$b_0c_0\tilde Q\Phi=0$.
The Siegel gauge and the Landau gauge are interpolated by one real parameter, called the $a$-gauge \cite{Asano:2006hk}.
}
\begin{align}
&{\cal P}_1=-\Bigl(c_0+\frac{\tilde Q}{L_0}\Bigr)b_0c_0W_1\tilde Q,
&{\cal P}_2=\Bigl(c_0+\frac{\tilde Q}{L_0}\Bigr)\Bigl(
b_0\Bigl(1+\frac{1}{L_0}\tilde QW_1\tilde Q\Bigr)-b_0c_0\tilde QW_1\Bigr),
\end{align}
where $\tilde Q$ is given by ghost zero mode expansion of $Q_{\rm B}$:
\begin{align}
Q_{\rm B}=c_0L_0+b_0M+\tilde Q
\end{align}
and $W_1$ is defined by
\begin{align}
&W_1=\sum_{k=0}^{\infty}\frac{(-1)^k}{((k+1)!)^2}M^k(M^-)^{k+1},
&M^-\equiv -\sum_{k=1}^{\infty}\frac{1}{2 k}b_{-k}b_k.
\end{align}

\subsection{On the level truncation method
\label{LTx}
}

In order to perform numerical calculations, we restrict ourselves to the
subspace spanned by the following basis.
\begin{itemize}
 \item We consider only the zero momentum sector with the ghost number 1.
 \item In the matter sector, except for the 25th direction, 
we use only the Virasoro generator with the central charge $c=25$, 
denoted as $L_{-n}^{({\rm m})\prime}$ $(n>1)$.
 \item As for the 25th sector, we use the conventional oscillator
       $\alpha_{-n}^{25}$ $(n\ge 1)$.
 \item In the ghost sector, we use $b_{-n}$ ($n>0$), $c_{-n}$ ($n\ge 0$)
on $c_1|0\rangle$, where $|0\rangle$ is the conformal vacuum.
 \item We take the even $\Omega'$ sector with $\Omega'\equiv
       (-1)^{L_0+1}P_{25}$ \cite{Sen:2000hx}, where $P_{25}$
       is a parity transformation with respect to the 25th direction 
such as $P_{25}\alpha_{-n}^{25}(P_{25})^{-1}=-\alpha_{-n}^{25}$,
       $P_{25}|0\rangle=|0\rangle$.
\end{itemize}
Using the above conditions, a general form of the basis is
\begin{align}
&L_{-n_1}^{({\rm m})\prime}L_{-n_2}^{({\rm m})\prime}
\cdots L_{-n_l}^{({\rm m})\prime}
\alpha^{25}_{-m_1}\alpha^{25}_{-m_2}\cdots \alpha^{25}_{-m_a}
b_{-k_1}b_{-k_2}\cdots b_{-k_b}
c_{-l_1}c_{-l_2}\cdots c_{-l_b}c_1|0\rangle,
\\
&n_1\ge \cdots n_l\ge 2,~~m_1\ge\cdots m_a\ge 1,~~k_1>\cdots k_b\ge 1,~~l_1>\cdots l_b\ge 0,\\
&n_1+\cdots n_l+m_1+\cdots m_a+k_1+\cdots k_b+l_1+\cdots l_b+a={\rm even}.
\end{align}
In fact, a space spanned by the above basis is closed under the action
of the operator (\ref{Q'x})  
and the star product and it is consistent with the Siegel and 
Landau gauge condition.

Furthermore, we use the $(L,3L)$-truncation method with respect to the level
associated with $L_0$. Namely, string fields are truncated
up to the level $L$, which is an eigenvalue of $L_0+1$, 
and each term of the expansion of the star product of string fields is
truncated up to the total level $3L$. 

Concretely, the dimension of the truncated space as above is $N_L$ 
in Table \ref{tab1} and $M_L$ is
that of the space where the Siegel or Landau gauge condition is imposed.
\begin{table}[H]
\begin{center}
\begin{tabular}{|c||c|c|c|c|c|c|c|c|c|}
\hline
$L$&0&1&2&3&4&5&6&7&8\\
\hline
$N_L$&1&2&6&12&29&56&118&218&420
\\
\hline
$M_L$&1&2&5&9&20&37&75&135&255
\\
\hline
\hline
$L$&9&10&11&12&13&14&15&16&$\cdots$\\
\hline
$N_L$&745&1348&2307&3985&6614&11011&17799&28764&$\cdots$
\\
\hline
$M_L$&446&797&1351&2315&3817&6317&10161&16346&$\cdots$
\\
\hline
\end{tabular}
\end{center}
\caption{
Dimensions of the truncated space for the level $L$. 
\label{tab1}
}
\end{table}
To solve (\ref{G-cond}), (\ref{P2EOM}) numerically, we use Newton's method.
With an appropriate initial configuration $\Phi_{(0)}$, we solve 
a set of linear equations:
\begin{align}
&{\cal P}_1\Phi_{(n+1)}=0,
\label{G-condit}\\
&{\cal P}_2(Q'\Phi_{(n+1)}+\Phi_{(n)}*\Phi_{(n+1)}+\Phi_{(n+1)}*\Phi_{(n)})={\cal P}_2(\Phi_{(n)}*\Phi_{(n)}),
\label{P2EOMit}
\end{align}
iteratively in the truncated space.
If $\lim_{n\to \infty}\Phi_{(n)}$ exists, it gives a solution to  (\ref{G-cond}) and (\ref{P2EOM}).
Actually, for a fixed truncation level $L$, 
we terminate the iterative procedure if the relative error of the convergence reaches
$\|\Phi_{(n+1)}-\Phi_{(n)}\|/\|\Phi_{(n)}\|<10^{-8}$.

We construct the tachyon vacuum solutions in the Siegel
gauge and the Landau gauge in the theory with $Q'$ (\ref{Q'x})
as follows:
\begin{itemize}
\item We begin by constructing a solution in the case of $x=0$ (the original theory with $Q_{\rm B}$).
We take $\Phi_{(0)}=\frac{64}{81\sqrt{3}}c_1|0\rangle$,
which is a nontrivial solution in the lowest level truncation,
 as an initial configuration
and then we get a converged solution $\Phi_{x=0}$,
which is twist even,
using the iterative procedure  (\ref{G-condit}) and (\ref{P2EOMit}).

\item In the case of a positive
value of $x$, we use a converged configuration $\Phi_{x-\epsilon}$
in the theory of $Q'$ with $x - \epsilon$ for a small value of $\epsilon(>0)$
as an initial configuration.
Solving  (\ref{G-condit}) and (\ref{P2EOMit}) iteratively, we get a converged solution 
in the theory of $Q'$ with $x$.

\item In the case of a negative value of $x$, noting (\ref{Q'x}),
 a numerical solution can be obtained by 
the parity transformation with respect to the 25th direction from the solution
in the theory of $Q'$ with $-x$, namely, $\Phi_x=P_{25}\Phi_{-x}$.

\item At large values of $|x|$, we reach the trivial solution $\Phi_x=0$ using the above procedure.

\end{itemize}
It turns out that it takes 10 iterations or fewer to get each converged solution.

\subsection{Evaluation of the vacuum energy
\label{sec:EVE}}

Here we demonstrate the results of the evaluation of
 the vacuum energy $E=V[\Phi_x]$ for the numerical solutions $\Phi_x$ obtained as in \S \ref{LTx}.
We normalize the potential  $V[\Phi]$ by a D-brane tension as
\begin{align}
V[\Phi]=-2\pi^2 S'[\Phi]=
2\pi^2\left(\frac{1}{2}\langle\Phi ,Q'\Phi\rangle\!+\!\frac{1}{3}\langle\Phi,\Phi*\Phi\rangle
\right).
\label{EPhi}
\end{align}
Noting the relation $V[\Phi_x]=V[P_{25}\Phi_{-x}]=V[\Phi_{-x}]$,
we consider only the case of nonnegative values of $x$.

\begin{figure}[t]
\begin{center}
\includegraphics[width = 9cm]{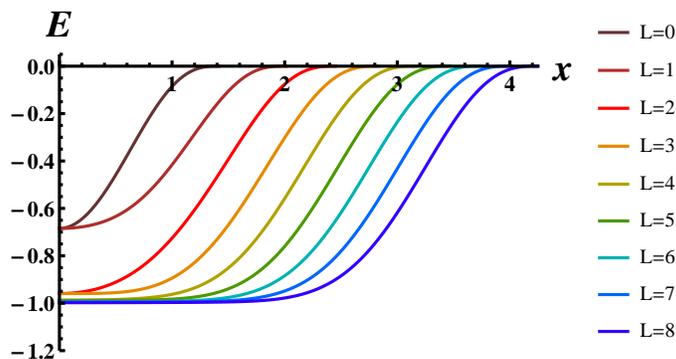}
\caption{
Plots of the vacuum energy $E=V[\Phi_x]$ (\ref{EPhi}) in the Siegel gauge for $L=0,1,2,\cdots,8$
truncation.
\label{fig:Sx}
}
\end{center}
\end{figure}

In the case of the Siegel gauge, we have Fig.~\ref{fig:Sx}.
For a fixed value of $x$, $E$ approaches $-1$ with increasing truncation level
and the region where $E\simeq -1$ becomes larger for higher levels.
In the infinite level limit, it seems to be $E=-1$ for all values of $x$.
Therefore, it is consistent that the numerical solutions $\Phi_x$
can be interpreted to represent the tachyon vacuum 
in the theory of $Q'$ with $x$, where a D-brane vanishes.

\begin{figure}[h]
\begin{center}
\includegraphics[width = 9cm]{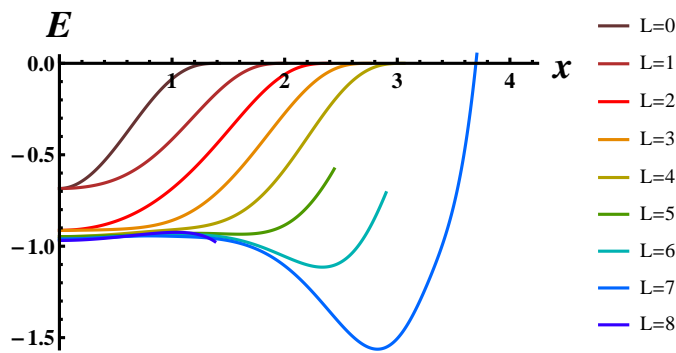}
\caption{
Plots of the vacuum energy $E=V[\Phi_x]$ (\ref{EPhi}) in the Landau
 gauge for $L=0,1,2,\cdots,8$ 
truncation.
\label{fig:Lx}
}
\end{center}
\end{figure}

In the case of the Landau gauge, we have Fig.~\ref{fig:Lx}.\footnote{
At $x=0$, the values of $E$ are the same as in \cite{Asano:2006hm} for
$L=0,2,4,6$ 
and they are slightly different from those in \cite{Kishimoto:2009cz} because 
a different projection for (\ref{P2EOM}) is adopted.
}
For a fixed value of $x$, $E$ approaches $-1$ with increasing level up to
$L=4$. However, for $L=5,6,7,8$, we cannot find
converged solutions for large $|x|$ with the same algorithm and the
value of $E$ seems to be unstable for $|x|>2$ even if there exist
numerical solutions. Compared to the result in the Landau gauge, the level
truncation in the Siegel gauge might be suitable for large values of
$|x|$. 

\subsection{Evaluation of the gauge invariant overlaps
\label{sec:EGIO}}

Here, we evaluate gauge invariant overlaps with the graviton and the
closed tachyon for the numerical solutions $\Phi_x$ obtained as in \S
\ref{LTx}. In general, the gauge invariant overlap $O_V(\Phi)$ is
defined as 
\begin{align}
O_V(\Phi)=\langle I|V(i)|\Phi\rangle.
\label{OVPhi}
\end{align}
Here, $V(i)$ is given by $c\bar cV_{\rm m}(z,\bar{z})$,
where $V_{\rm m}(z,\bar z)$ is a vertex operator in the matter sector
with the conformal dimension $(1,1)$.
(See \cite{Kawano:2008ry} for details of explicit calculations.)

We evaluate the gauge invariant overlap with
the graviton: $V_{\rm m}\sim \partial X^0(i)\partial X^0(-i)$,
where we denote (\ref{OVPhi}) as $O_{\zeta}(\Phi)$,
and the closed tachyon $V_{\rm m}\sim e^{\frac{i}{2}k
(X^{25}(i)-X^{25}(-i))}$ with $k^2=4/\alpha'$
for a Dirichlet direction, where we denote  (\ref{OVPhi}) as
$O_k(\Phi)$. 
We normalize them as 
\begin{align}
O_{\zeta}(\Phi_T)&=1,
\label{OzetaPhiT}
\\
O_k(\Phi_T)&=e^{-4i x}
\label{OkPhiT}
\end{align}
for the analytic solution (\ref{PhiT}) 
using the result in \cite{Inatomi:2012nv, Inatomi:2012nd}.

\begin{figure}[h]
\begin{center}
\includegraphics[width = 10.5cm]{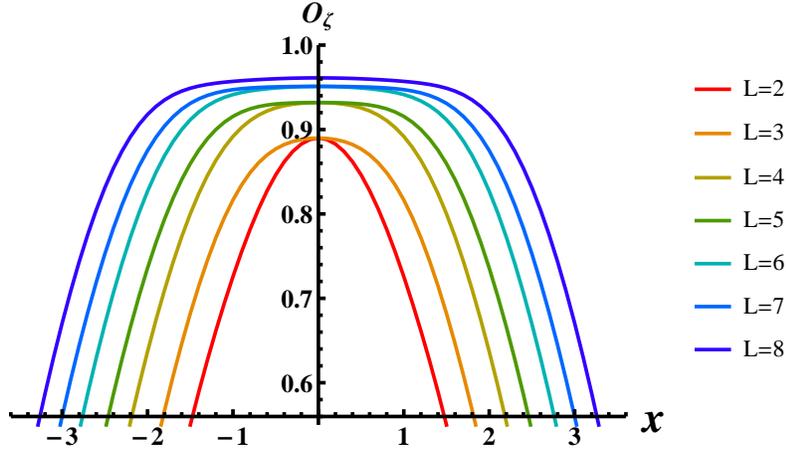}
\caption{
Plots of the gauge invariant overlap (\ref{OVPhi}) with the graviton
 $O_{\zeta}(\Phi_x)$ in the Siegel gauge for the tachyon vacuum using
 $L=2,3,\cdots,8$ truncation. 
\label{fig:GIO00x}
}
\end{center}
\end{figure}

In the Siegel gauge, we have evaluated 
the gauge invariant overlap with the graviton for 
the tachyon vacuum solution in the theory of $Q'$ with $x$
as in Fig.~\ref{fig:GIO00x}.
With increasing level, it 
approaches a constant near $1$ 
for larger regions of $x$, 
which is the same value as in  (\ref{OzetaPhiT}).
Namely, in the infinite level limit,
we expect $O_{\zeta}(\Phi_x)=1$
for all $x$.

\begin{figure}[h]
\begin{center}
\includegraphics[width = 10.5cm]{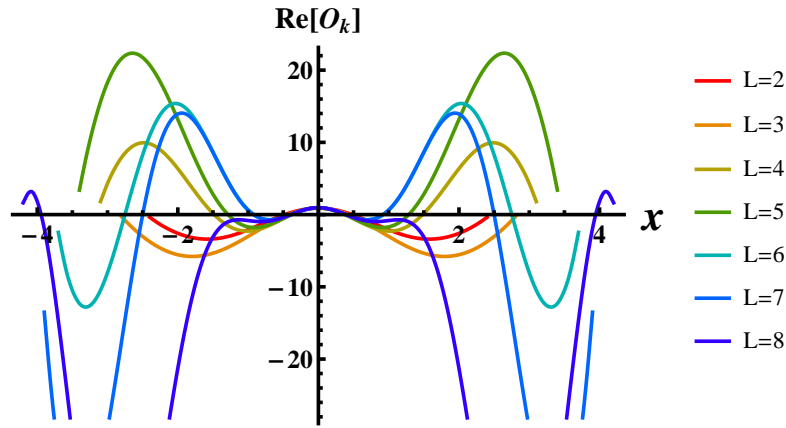}
\caption{
Plots of the real part of the gauge invariant overlap (\ref{OVPhi}) with the closed tachyon 
 $O_k(\Phi_x)$  in the Siegel gauge for 
the tachyon vacuum using $L=2,3,\cdots,8$ truncation.
\label{fig:GIO25xr}
}
\end{center}
\end{figure}

\begin{figure}[h]
\begin{center}
\includegraphics[width = 10.5cm]{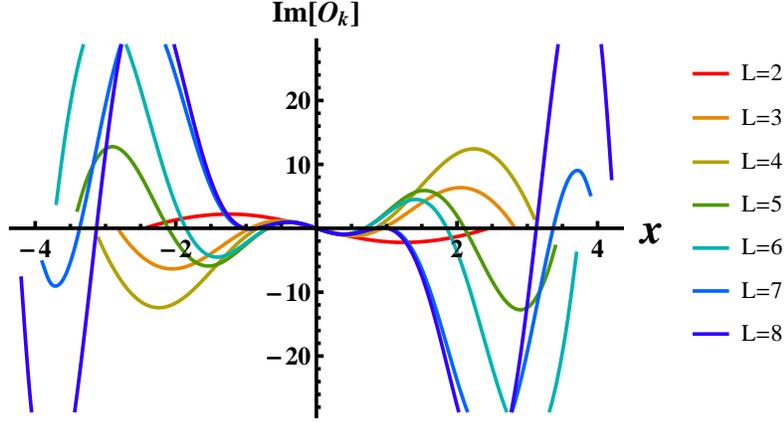}
\caption{
Plots of the imaginary part of the gauge invariant overlap (\ref{OVPhi}) with the closed tachyon 
 $O_k(\Phi_x)$  in the Siegel gauge for 
the tachyon vacuum using $L=2,3,\cdots,8$ truncation.
\label{fig:GIO25xi}
}
\end{center}
\end{figure}

The gauge invariant overlap with the closed tachyon for 
the tachyon vacuum solution in the theory of $Q'$ with $x$
is evaluated
as in Fig.~\ref{fig:GIO25xr}
for its real part, and Fig.~\ref{fig:GIO25xi}
for its imaginary part.

\begin{figure}[h]
\begin{center}
\includegraphics[width = 10.5cm]{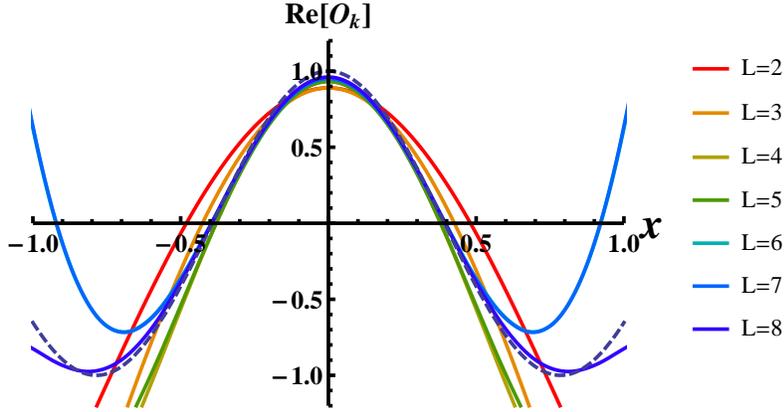}
\caption{
Superposition of Fig.~\ref{fig:GIO25xr} and $\cos 4x$ (dotted line)
for small values of $|x|$.
\label{fig:GIO25xr_scos}
}
\end{center}
\end{figure}

\begin{figure}[h]
\begin{center}
\includegraphics[width = 10.5cm]{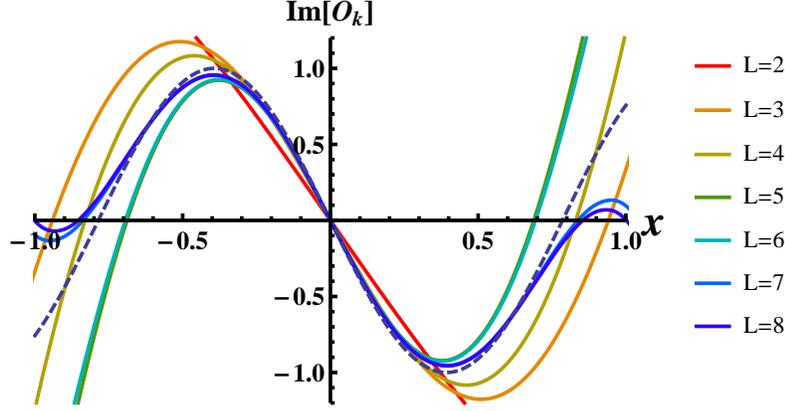}
\caption{
Superposition of Fig.~\ref{fig:GIO25xi} and $-\sin 4x$ (dotted line)
for small values of $|x|$.
\label{fig:GIO25xi_ssin}
}
\end{center}
\end{figure}

At first sight, the plots of $O_k(\Phi_x)$
 in Figs.~\ref{fig:GIO25xr} and \ref{fig:GIO25xi}
seem to be divergent for higher levels 
especially for large values of $|x|$.
However, for small values of $|x|$,
one can expect a structure like (\ref{OkPhiT}).
Actually, it seems to become  $O_k(\Phi_x)\to e^{-4ix}$
with increasing truncation level,
at least for small values of $|x|$
as in Figs.~\ref{fig:GIO25xr_scos} and \ref{fig:GIO25xi_ssin}.

\section{$M$-branch and $V$-branch in the $Q'$-theory
\label{sec:MV}
}

In this section, we investigate the $M$-branch 
and the $V$-branch in the theory
of $Q'$ with $x$. 
Firstly, we consider them in the lowest level.
In the $L=1$ truncation, the string field $\Phi$ is expressed as
\begin{align}
\Phi_{L=1}=t_0c_1|0\rangle+a_s\alpha^{25}_{-1}c_1|0\rangle.
\end{align}
Substituting it into the action (\ref{S'Phi}), we have
\begin{align}
S'[\Phi_{L=1}]=\frac{t_0^2}{2}-\frac{x^2t_0^2}{4}-\frac{x^2a_s^2}{4}-\frac{x a_s t_0}{\sqrt{2}}
-\frac{27\sqrt{3}}{64}t_0^3-\frac{3\sqrt{3}}{4}a_s^2t_0.
\end{align}
Solving $\frac{\partial}{\partial t_0}S'[\Phi_{L=1}]=0$
with respect to $t_0$, we have two solutions as functions of $a_s$:
\begin{align}
t_0^{(\mp)}(a_s)=\frac{4}{81\sqrt{3}}\left(
8-4x^2\mp \sqrt{
16(2-x^2)^2-162\sqrt{6} x a_s-729 a_s^2
}
\right).
\label{t0as}
\end{align}
One of them satisfies $t_0^{(-)}(a_s=0)=0$, which corresponds to the $M$-branch,
and another one satisfies  $t_0^{(+)}(a_s=0)\ne 0$, which corresponds to the $V$-branch
(for small values of $|x|$).
Both of them exist only in a finite interval:
\begin{align}
-\frac{\sqrt{2}}{27}\left(3\sqrt{3}x+\sqrt{32-5 x^2+8x^4}\right)\le a_s
\le \frac{\sqrt{2}}{27}\left(-3\sqrt{3}x+\sqrt{32-5 x^2+8x^4}\right)
\end{align}
because of the reality of the tachyon field $t_0$.
At the end of the interval, the two branches merge.

\begin{figure}[h]
\parbox{.3\textwidth}{
\begin{center}
\includegraphics[width = 5.5cm]{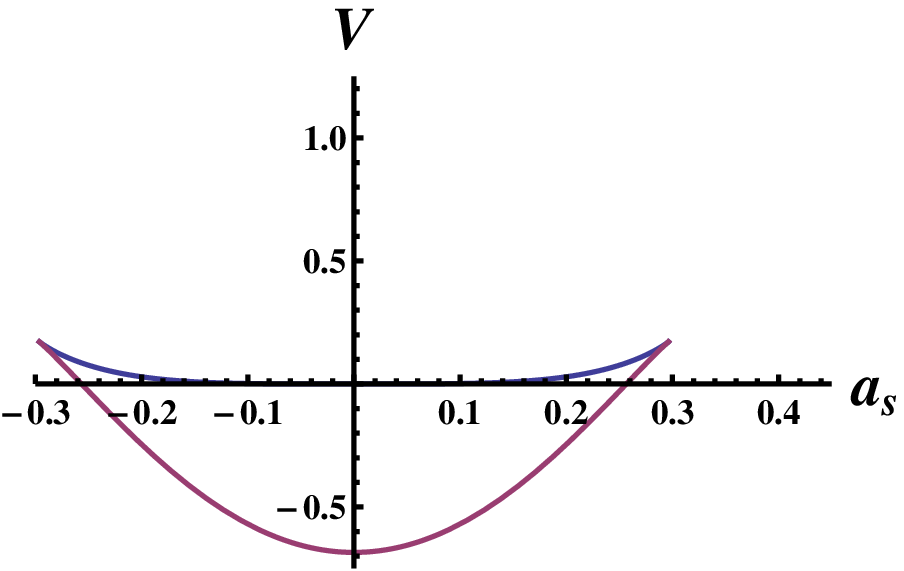}
\caption{The $M$-branch and the $V$-branch for $L=1$ in the original theory
 (i.e. $x=0$). 
\label{fig:L1x0MV}
}
\end{center}
}
\hfill
\parbox{.3\textwidth}{
\begin{center}
\includegraphics[width = 5.5cm]{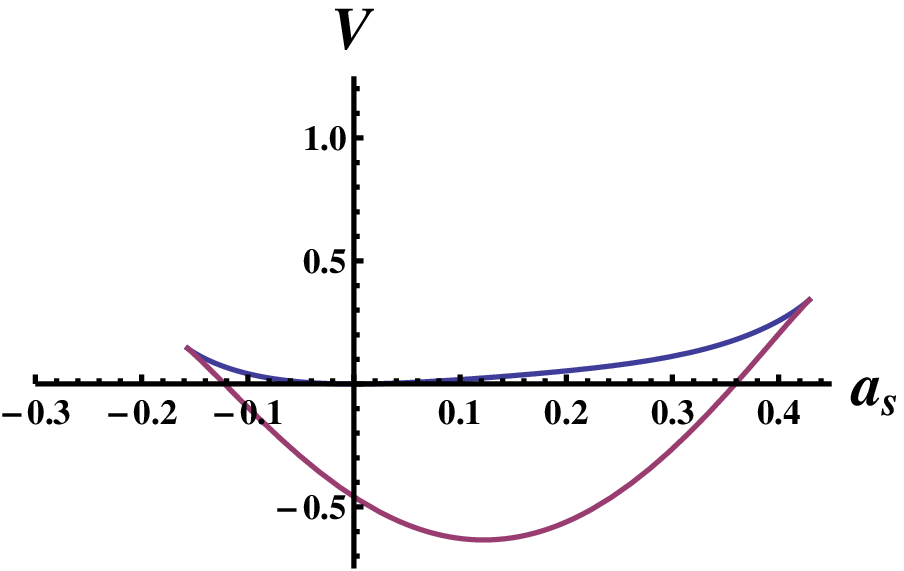}
\caption{
The $M$-branch and the $V$-branch for $L=1$ in the theory with $x=-0.5$.
\label{fig:L1xm0p5MV}
}
\end{center}
}
\hfill
\parbox{.3\textwidth}{
\begin{center}
\includegraphics[width = 5.5cm]{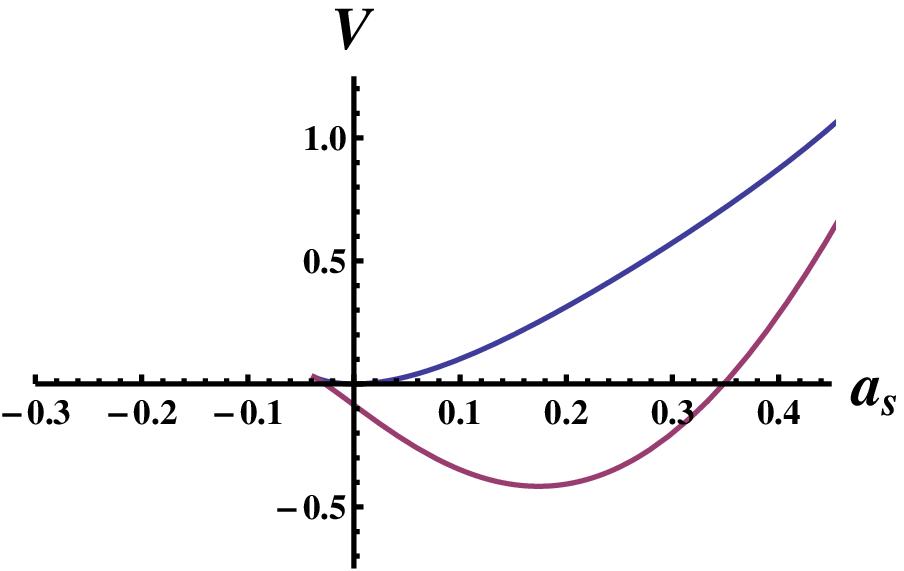}
\caption{
The $M$-branch and the $V$-branch for $L=1$ in the theory with $x=-1$.
\label{fig:L1xm1MV}
}
\end{center}
}
\end{figure}

\begin{figure}[h]
\parbox{.3\textwidth}{
\begin{center}
\includegraphics[width = 5.5cm]{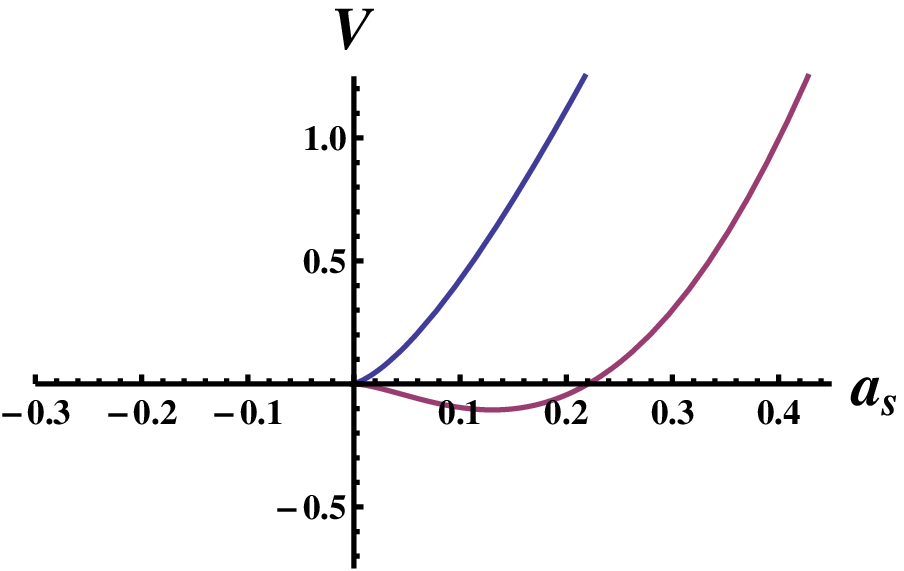}
\caption{
The $M$-branch and the $V$-branch for $L=1$ in the theory with $x=-1.5$.
\label{fig:L1xm1p5MV}
}
\end{center}
}
\hfill
\parbox{.3\textwidth}{
\begin{center}
\includegraphics[width = 5.5cm]{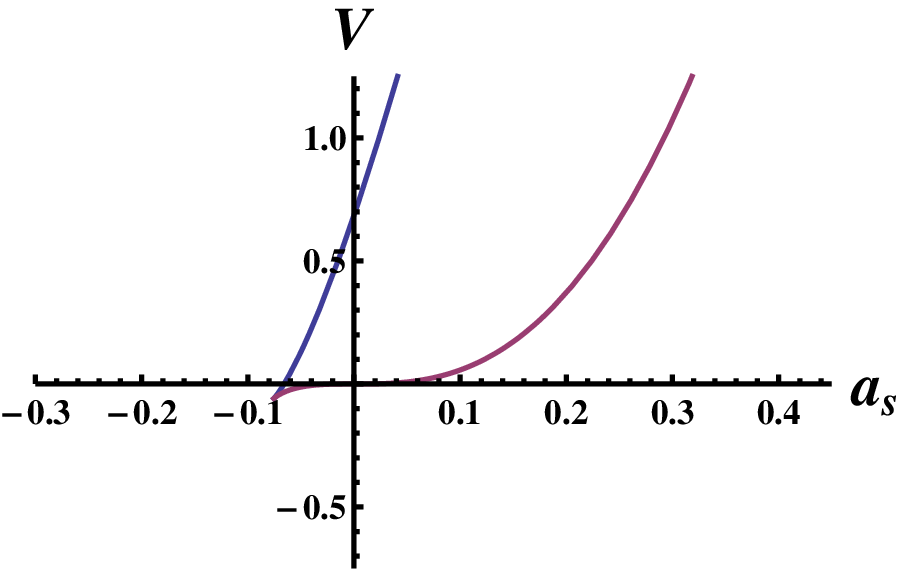}
\caption{
The $M$-branch and the $V$-branch for $L=1$ in the theory with $x=-2$.
\label{fig:L1xm2MV}
}
\end{center}
}
\hfill
\parbox{.3\textwidth}{
\begin{center}
\includegraphics[width = 5.5cm]{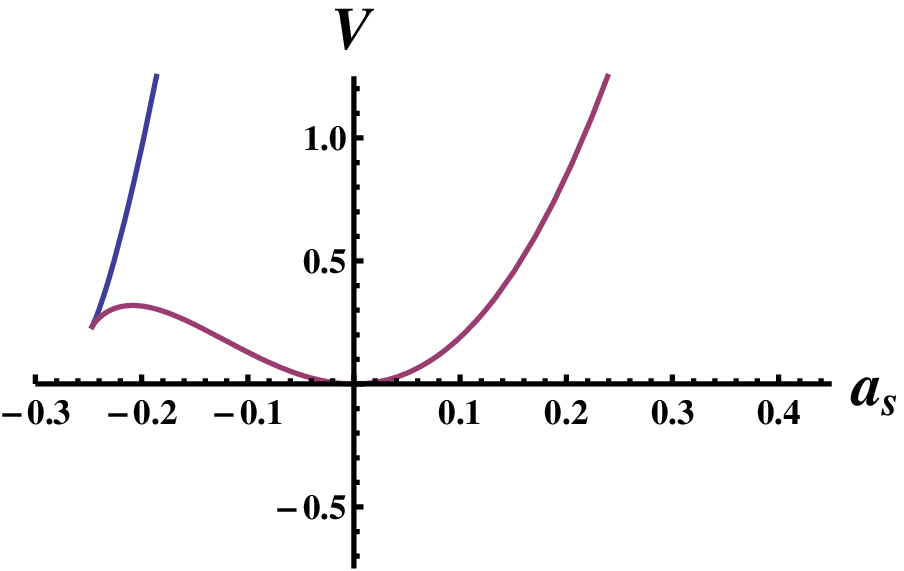}
\caption{
The $M$-branch and the $V$-branch for $L=1$ in the theory with $x=-2.5$.
\label{fig:L1xm2p5MV}
}
\end{center}
}
\end{figure}

Substituting $\Phi_{L=1}$ with these $t_0^{(\mp)}(a_s)$ (\ref{t0as}) in
$V[\Phi]$ (\ref{EPhi}), 
we obtain an effective potential as a function of $a_s$.
For example, we have Figs.~\ref{fig:L1x0MV}, \ref{fig:L1xm0p5MV},
\ref{fig:L1xm1MV}, 
 \ref{fig:L1xm1p5MV}, \ref{fig:L1xm2MV}, \ref{fig:L1xm2p5MV}
for the theory with $x=0,-0.5,-1,-1.5,-2,-2.5$, respectively.
The $M$-branch is depicted by $V_M(a_s)\equiv
V[\Phi_{L=1}|_{t_0=t_0^{(-)}(a_s)}]$ 
and the $V$-branch is depicted by $V_V(a_s)\equiv
V[\Phi_{L=1}|_{t_0=t_0^{(+)}(a_s)}]$. 
We find $V_M(a_s)\ge V_V(a_s)$ from explicit expressions.

In the case of $x^2<2$,  we have expansions
of two branches around $a_s=0$ as
\begin{align}
&V_M(a_s)=\frac{\pi^2x^2(-4+x^2)}{2(-2+x^2)}a_s^2+O(a_s^3),
&V_V(a_s)=\frac{512\pi^2(-2+x^2)^3}{59049}+O(a_s).
\end{align}
Similarly, for  $x^2>2$, we have
\begin{align}
&V_M(a_s)=\frac{512\pi^2(-2+x^2)^3}{59049}+O(a_s),
&V_V(a_s)=\frac{\pi^2x^2(-4+x^2)}{2(-2+x^2)}a_s^2+O(a_s^3).
\end{align}
Therefore, we note that $V_V(a_s)$ has a second-order zero at $a_s=0$
for $|x|>\sqrt{2}$.\footnote{
In the case of $|x|=\sqrt{2}$, both $V_M(a_s)$ and $V_V(a_s)$ 
become $O(a_s\sqrt{a_s})$ around $a_s=0$.
}
Actually, from Figs.~\ref{fig:L1xm2MV} and \ref{fig:L1xm2p5MV}, 
the graph of the $V$-branch for $|x|>\sqrt{2}$, given by $V_V(a_s)$,  
may be qualitatively similar to the ``$M$-branch'' in the original theory.
\\

For higher level truncation, we solve eqs. (\ref{G-cond}) and (\ref{P2EOM})
 with a fixed value of the massless field $a_s$
using the iterative method with appropriate initial configurations.
In a fixed level $L$ truncation, for a fixed value of $x$, we take initial configurations as follows:
\begin{itemize}
\item For the $V$-branch, we begin from the value of $a_s(\equiv a_s^T)$ of
the tachyon vacuum solution constructed as in \S\ref{LTx}.
Using the values of component fields of the tachyon vacuum except for $a_s$
as an initial configuration,
we solve eqs. (\ref{G-cond}) and (\ref{P2EOM})
for $a_s=a_s^T \pm \varepsilon$ for a small value of $\varepsilon(>0)$.
Then, we use the values of component fields of the converged solution
with $a_s=a_s^T \pm \varepsilon$,
      except for $a_s$, as an initial configuration
for the iteration with $a_s=a_s^T \pm 2 \varepsilon$.
Similarly, we use the configuration of the converged solution with
      $a_s=a_s^T \pm 2 \varepsilon$
to solve  (\ref{G-cond}) and (\ref{P2EOM})  with $a_s=a_s^T \pm 3
      \varepsilon$, and so on.

\item For the $M$-branch, 
we begin from the value of $\varepsilon$ ($-\varepsilon$)
for $a_s$ and we use zeros for values of component fields except for
      $a_s$
as an initial configuration to construct a solution of  
(\ref{G-cond}) and (\ref{P2EOM}).
Then, we use the values of the converged solution 
with $a_s= \varepsilon$  ($a_s= -\varepsilon$),
except for $a_s$,
as an initial configuration
for the iteration with $a_s= 2 \varepsilon$  ($a_s= -2 \varepsilon$).
Similarly, we use the configuration of  the converged solution with
      $a_s= \pm 2 \varepsilon$
to solve  (\ref{G-cond}) and (\ref{P2EOM})  with $a_s= \pm 3 \varepsilon$,
 and so on.

\end{itemize}
We consider numerical solutions only in the Siegel gauge
because it seems to be more stable than the Landau gauge,
as seen in \S \ref{sec:EVE}.
All component fields can be expressed as functions of $a_s$ numerically
and we substitute them to the potential (\ref{EPhi}) to get an effective
potential $V_{\rm S}$ as a function of $a_s$.

\begin{figure}[h]
\begin{center}
\includegraphics[width = 10.5cm]{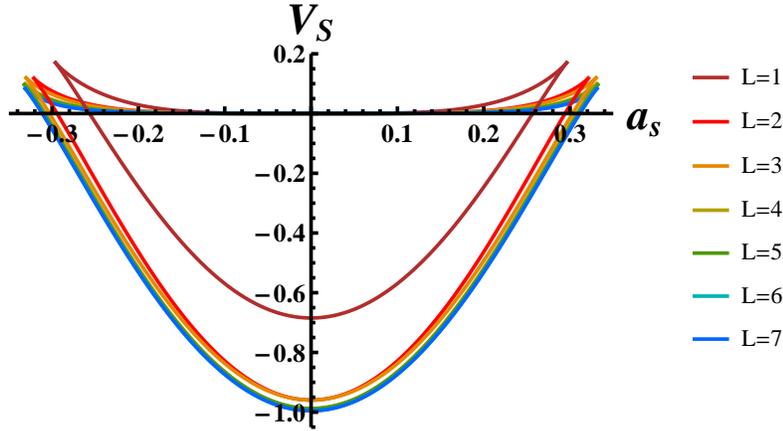}
\caption{
The $M$-branch and the $V$-branch in the Siegel gauge
in the original theory $(x=0)$
with level  $L=1,2,\cdots,7$ truncation. 
\label{fig:szL7}
}
\end{center}
\end{figure}

\begin{figure}[h]
\begin{center}
\includegraphics[width = 11cm]{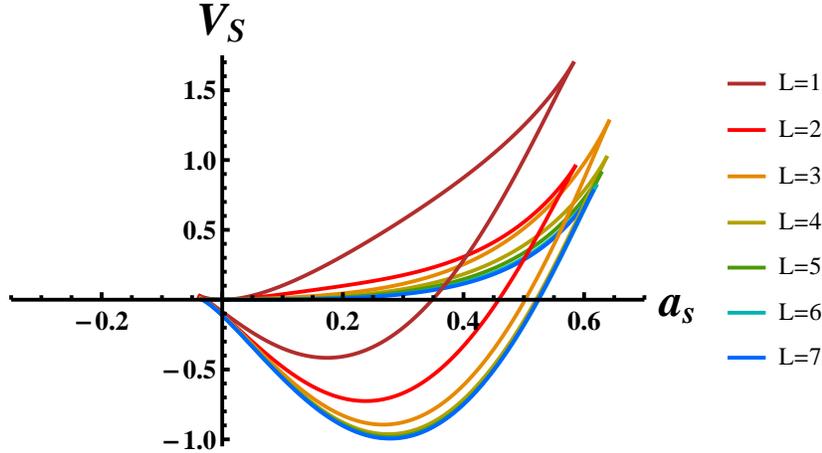}
\caption{
The $M$-branch and the $V$-branch in the Siegel gauge
in the theory of $x=-1$
with level  $L=1,2,\cdots,7$ truncation. 
\label{fig:szm1L7}
}
\end{center}
\end{figure}

In the original theory ($x=0$), we have computed 
the $M$-branch and the $V$-branch as shown in Fig.~\ref{fig:szL7},
which was already given in \cite{Sen:2000hx} up to level  $L=4$.\footnote{
Precisely speaking, the method of $(L,2L)$ approximation instead of
$(L,3L)$ was performed in \cite{Sen:2000hx}. 
}
In the theory of $Q'$ with $x=-1$, we find the $M$-branch and the
$V$-branch as shown in Fig.~\ref{fig:szm1L7}.
In both cases, with increasing level,
the plots of the $M$-branch become flatter and
it seems that there exist a maximum and minimum of the values of the
massless field $a_s$, where the $M$-branch and the $V$-branch merge, for
further higher levels.
Comparing  Fig.~\ref{fig:szm1L7} with  Fig.~\ref{fig:szL7}, 
the qualitative features of the graphs are similar,
except that both branches move 
in the horizontal direction.

\subsection{$M$- and $V$-branches for various values of $x$ in $L=6$}

Here, we demonstrate the numerical results in the level $L=6$ truncation.
 
\begin{figure}[h]
\begin{center}
\includegraphics[width = 9.5cm]{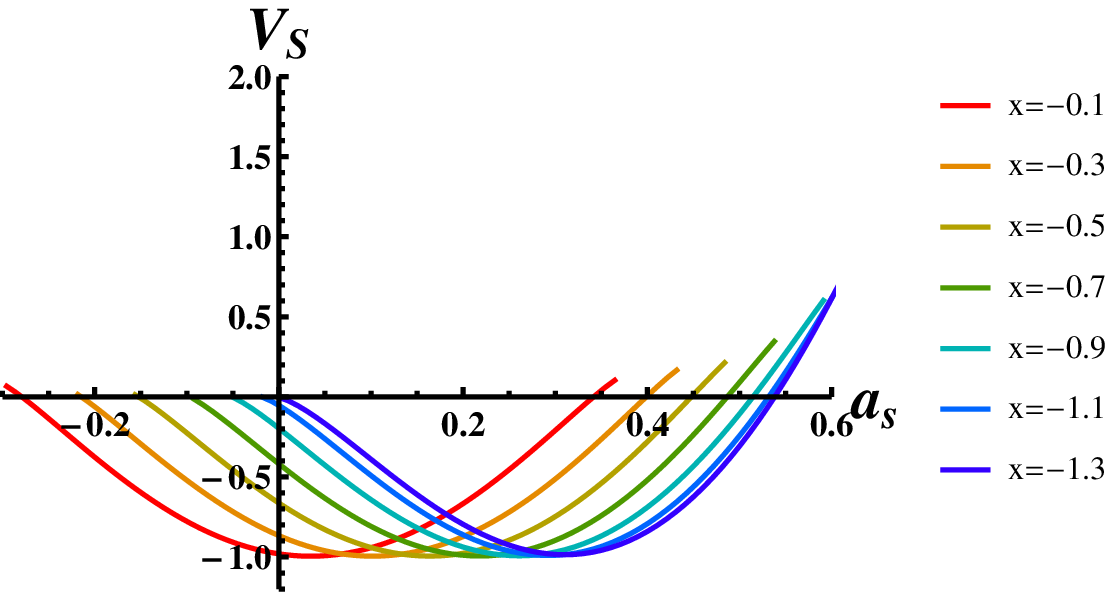}
\caption{
The $V$-branch in the Siegel gauge
in the theory of $Q'$ with
$x=-0.1,-0.3,-0.5,\cdots,-1.3$
with level  $L=6$ truncation. \sloppy
\label{fig:L6V0to1p3}
}
\end{center}
\end{figure}

\begin{figure}[h]
\begin{center}
\includegraphics[width = 11cm]{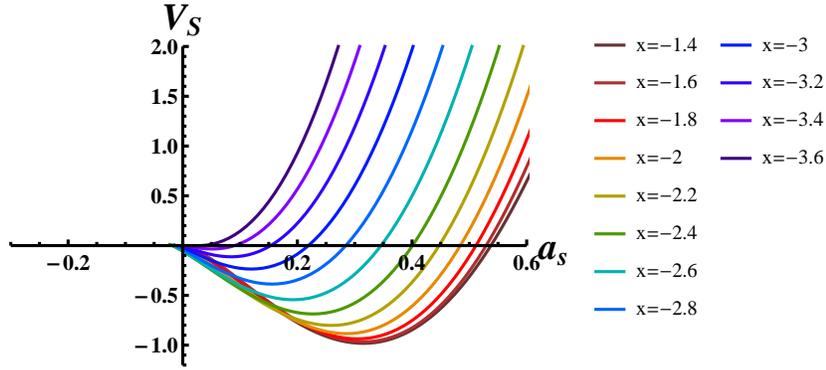}
\caption{
The $V$-branch in the Siegel gauge
in the theory of $Q'$ with $x=-1.4,-1.6,\cdots,-3.6$
with level  $L=6$ truncation. 
\label{fig:L6V1p4to3p7}
}
\end{center}
\end{figure}

For small values of  $|x|$, we have plots of $V$-branches as in
Fig.~\ref{fig:L6V0to1p3} and for large values of $|x|$, we have those in
Fig.~\ref{fig:L6V1p4to3p7}.
In Fig.~\ref{fig:L6V0to1p3}, the $V$-branch plot moves to
the right in the horizontal direction
when the value of $x$ changes from $x=0$ to $x=-1.3$.
Then, in Fig.~\ref{fig:L6V1p4to3p7},
the left end of the $V$-branch remains near the origin
(i.e. $a_s=0$)
and the potential minimum 
moves to the upper left
when the value of $x$ changes from $x=-1.4$ to $x=-3.6$.
In the end, i.e. $x\simeq -3.6$, the $V$-branch 
is similar to the ``$M$-branch''
in the sense that the plot appears to be  flat
around $a_s\gtrsim 0$.\\

\begin{figure}[h]
\begin{center}
\includegraphics[width = 9.5cm]{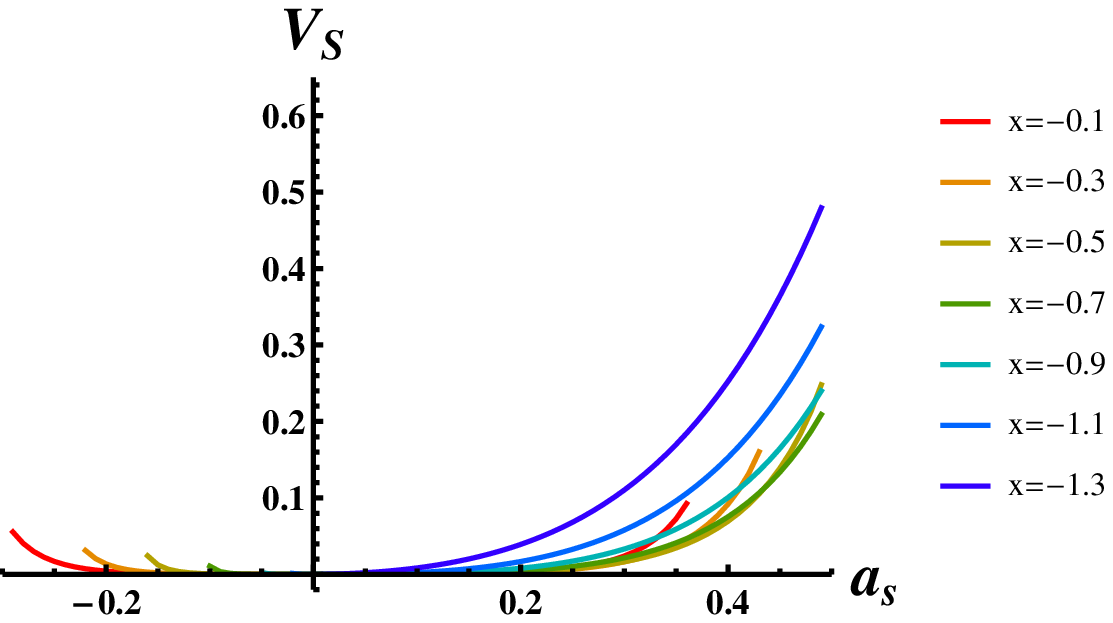}
\caption{
The $M$-branch in the Siegel gauge
in the theory of $Q'$ with $x=-0.1,-0.3,-0.5,\cdots,-1.3$
with level $L=6$ truncation. 
\label{fig:L6M0to1p4}
}
\end{center}
\end{figure}

\begin{figure}[h]
 \begin{center}
\includegraphics[width = 10.5cm]{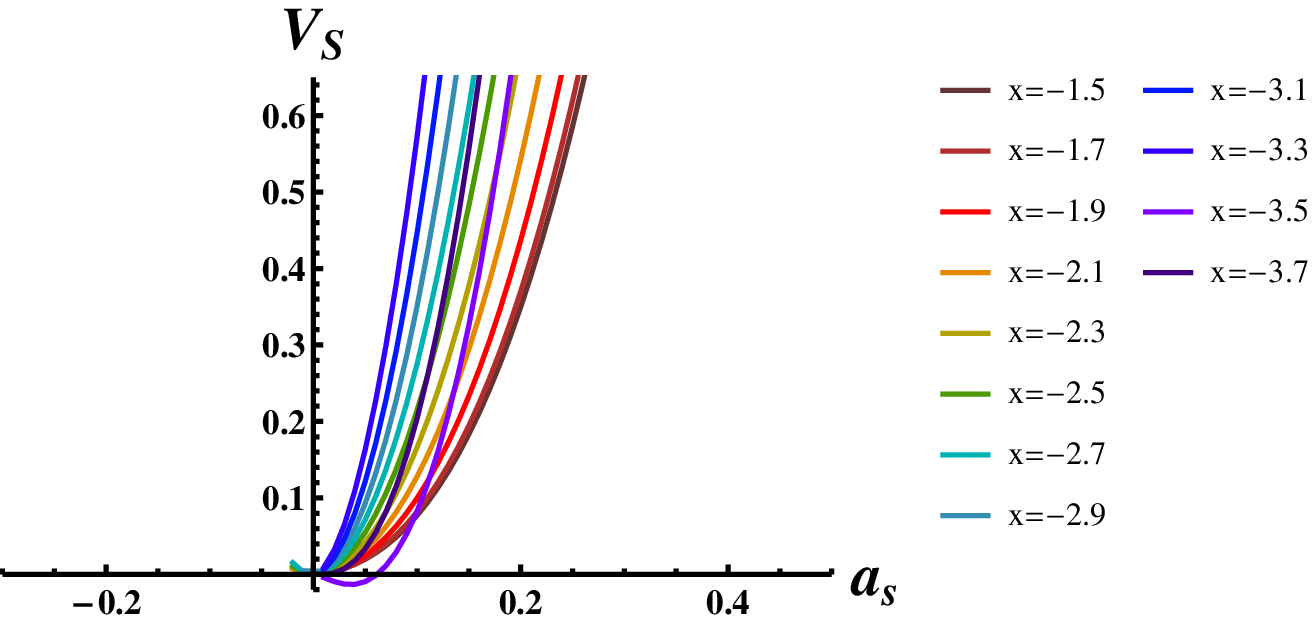}
\caption{
The $M$-branch in the Siegel gauge
in the theory of $Q'$ with $x=-1.5,-1.7,\cdots,-3.7$
with level $L=6$ truncation. 
\label{fig:L6M1p5to3p7}
}
\end{center}
\end{figure}

On the other hand, 
for small values of  $|x|$, we have plots of the $M$-branches as in
Fig.~\ref{fig:L6M0to1p4} and for large values of $|x|$, we have those in 
Fig.~\ref{fig:L6M1p5to3p7}.
In Fig.~\ref{fig:L6M0to1p4}, the $M$-branch plot moves to
the right when the value of $x$ changes from $x=0$ to $x=-1.3$.
Then, in Fig.~\ref{fig:L6M1p5to3p7},
the left end of the $V$-branch remains near the origin
and the value of the potential suddenly increases for positive 
values of $a_s$ when the value of $x$ changes from $x=-1.5$ to $x=-3.7$.
In this sense, the $M$-branch seems to be unstable for $x < -1.4$.

\subsection{On a bound of $|a_s|$}

\begin{figure}[h]
\begin{center}
\includegraphics[width=9.5cm]{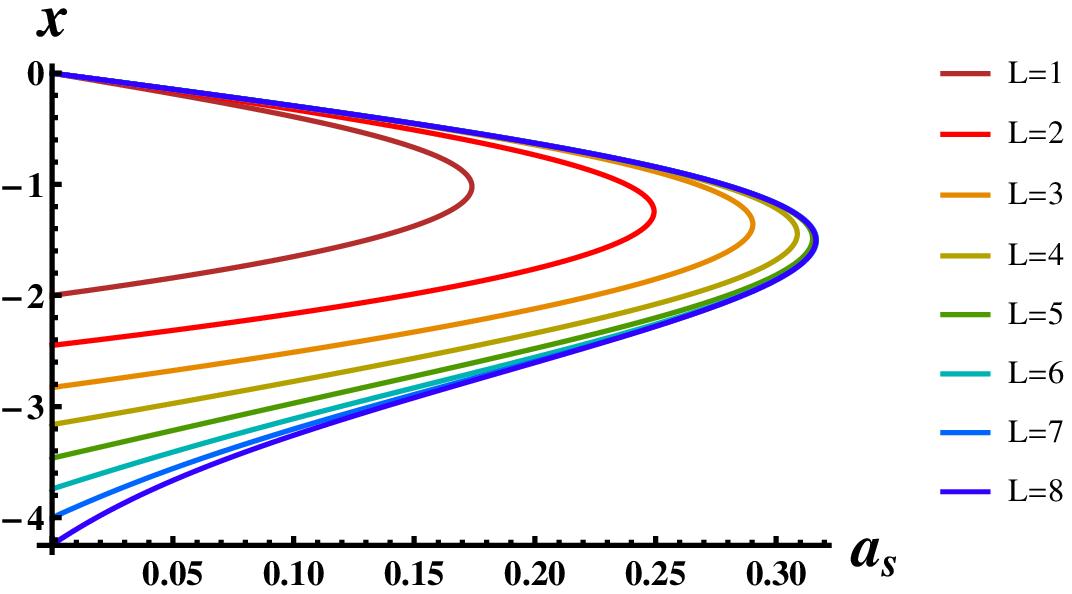}
\caption{
Plots of $a_s$ as a function of $x$ for the tachyon vacuum in the Siegel gauge
 using the numerical results in the level $L=1,2,\cdots,8$ truncation.
\label{as_x_comp}}
\end{center}
\end{figure}

{}From the results so far, there seems to be a finite bound 
on the value of the massless field $a_s$ for the numerical solutions 
in the Siegel gauge even in the theory of $Q'$ with $x\ne 0$.
Let us investigate the $x$-dependence
 of the value of  $a_s$ at the tachyon vacuum,
which is the minimum in the $V$-branch,
in the Siegel gauge.
We have obtained the numerical result shown in Fig.~\ref{as_x_comp}.
{}From Fig.~\ref{as_x_comp}, the plot seems to be convergent
to a curve, which has a finite maximum of $a_s$, 
in the limit $L\to \infty$.

\begin{figure}[h]
\parbox{.3\textwidth}{
\begin{center}
\includegraphics[width=5.1cm]{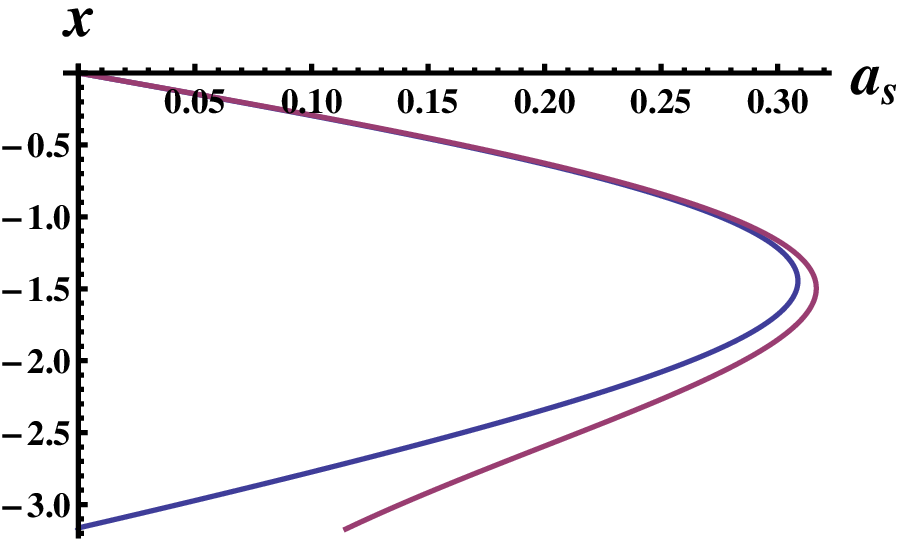}
\caption{
Plots of Fig.~\ref{as_x_comp} and $a_s$ given by (\ref{as_infty})
using numerical data in the level $L=4$ truncation.
\label{fig:L4as_x}}
\end{center}
}
\hfill
\parbox{.3\textwidth}{
\begin{center}
\includegraphics[width=5.1cm]{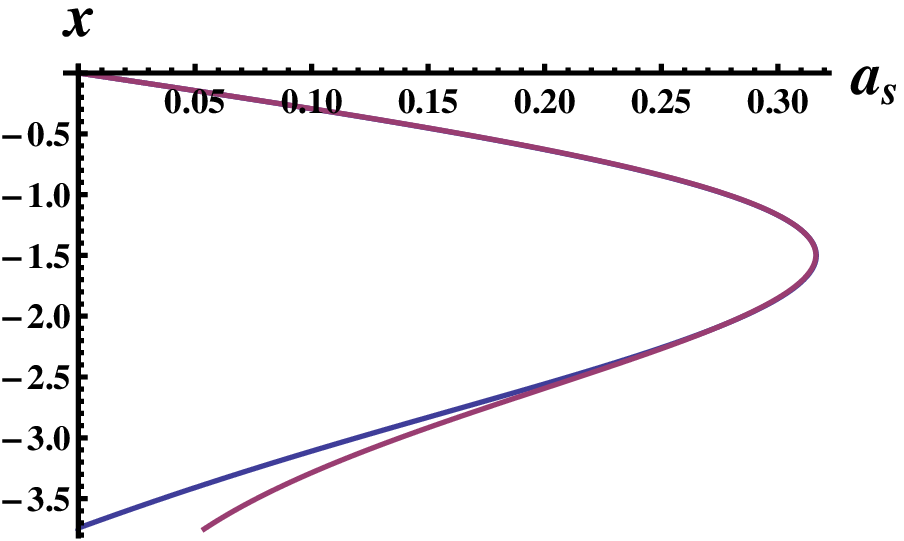}
\caption{
Plots of Fig.~\ref{as_x_comp} and $a_s$ given by (\ref{as_infty})
using numerical data in the level $L=6$ truncation.
\label{fig:L6as_x}}
\end{center}
}
\hfill
\parbox{.3\textwidth}{
\begin{center}
\includegraphics[width=5.1cm]{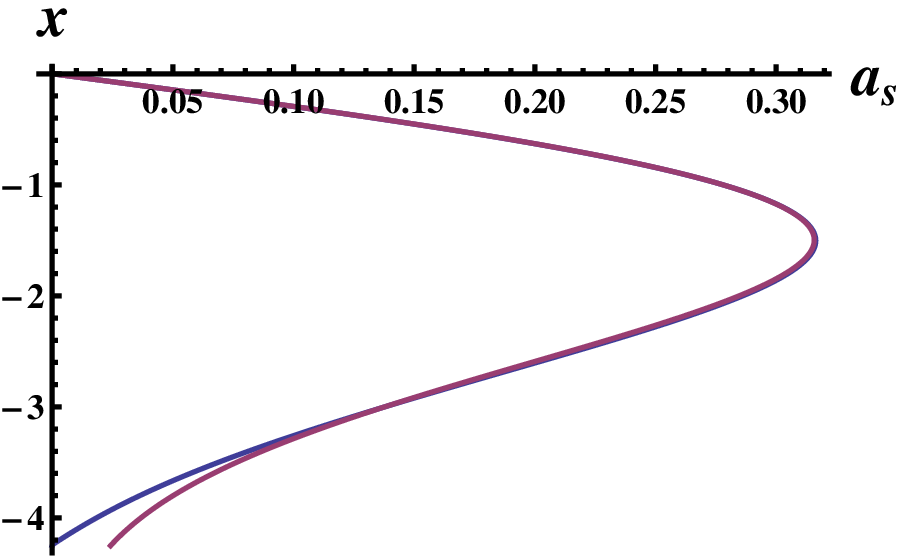}
\caption{
Plots of Fig.~\ref{as_x_comp} and $a_s$ given by (\ref{as_infty})
using numerical data in the level $L=8$ truncation.
\label{fig:L8as_x}}
\end{center}
}
\end{figure}

\begin{figure}[h]
\parbox{.471\textwidth}{
\begin{center}
\includegraphics[width=6.1cm]{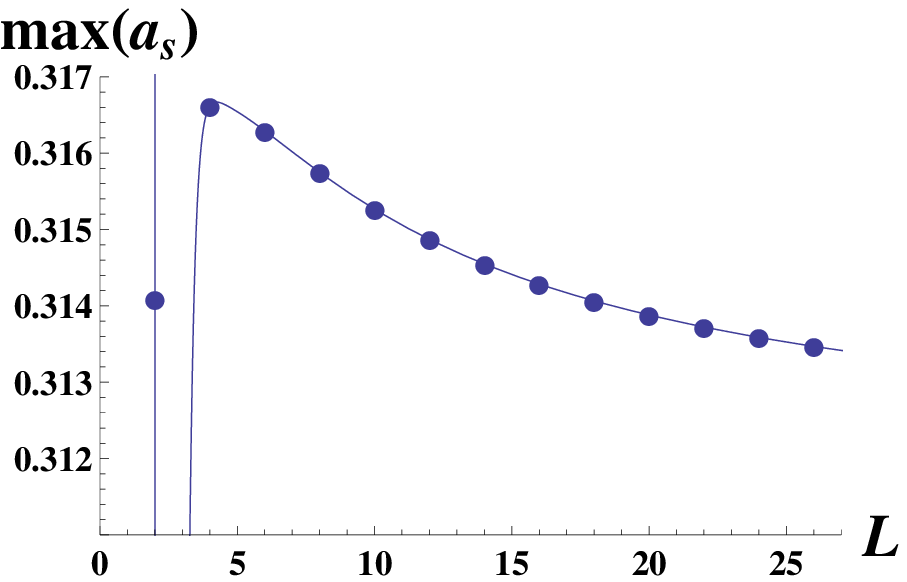}
\caption{The maximum of $a_s$ (\ref{as_infty}) using data in the
 truncation level $L=2,4,\cdots,26$
in the original theory.
The solid line denotes the plot of a fitting function of the form $\sum_{k=0}^{13}c_k/L^k$.
\label{fig:maxas_L}}
\end{center}
}
\hfill
\parbox{.471\textwidth}{
\begin{center}
\includegraphics[width=6.1cm]{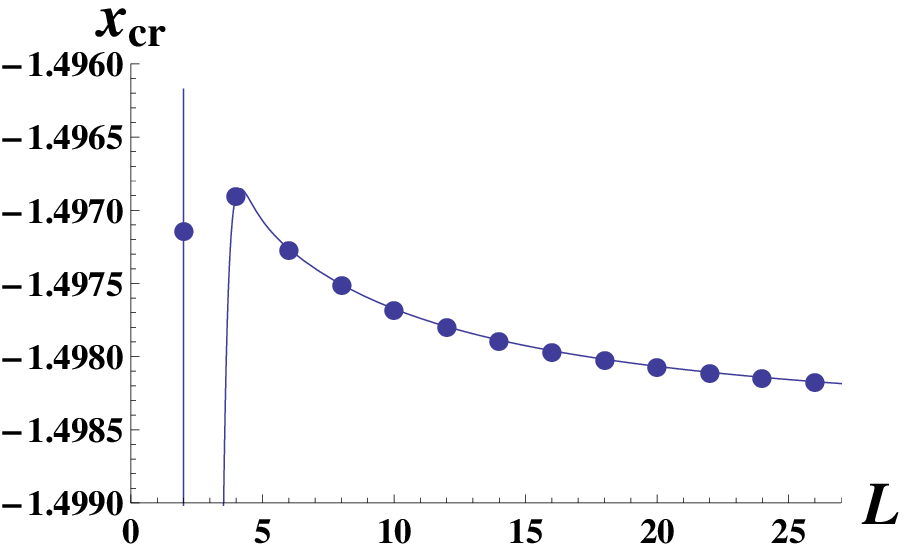}
\caption{
Plot of $x_{\rm cr}$ giving $\max(a_s)$ as in Fig.~\ref{fig:maxas_L}.
The solid line denotes the plot of fitting function of the form $\sum_{k=0}^{13}c_k/L^k$.
\label{fig:xcr_L}}
\end{center}
}
\end{figure}

Actually, we can explain this $x$-dependence of  $a_s$ 
in the Siegel gauge as follows.
We note that (\ref{G-cond}) and (\ref{P2EOM}) can be rewritten as
\begin{align}
&b_0\Phi=0,
&L'\Phi+b_0(\Phi*\Phi)=0,
\end{align}
where $L'$ can be expressed, using $U(x)$ such as $U(x)^{\dagger}=U(x)^{-1}$,
as follows:
\begin{align}
L'&\equiv\{b_0,Q'\}
=L_0+\frac{x}{\sqrt{2}}(\alpha_{-1}^{25}+\alpha_1^{25})+\frac{x^2}{2}
=U(x)L_0U(x)^{-1},
\label{L'def}
\\
U(x)&=\exp\left(\frac{x}{\sqrt{2}}(\alpha_1^{25}-\alpha_{-1}^{25})\right)
=e^{-\frac{1}{4}x^2}\exp\left(-\frac{x}{\sqrt{2}}\alpha_{-1}^{25}\right)
\exp\left(\frac{x}{\sqrt{2}}\alpha_1^{25}\right).
\end{align}
For any string fields $A,B$, we have
\begin{align}
U(x)^{-1}(A*B)=(U(x)^{-1}A)*(U(x)^{-1}B)
\label{UxinA*B}
\end{align}
and therefore the tachyon vacuum solutions $\Phi_x$ in
the Siegel gauge in the theory of $Q'$ 
with different values of $x$ 
can be related as
\begin{align}
\Phi_{x_2}=U(x_2)U(x_1)^{-1}\Phi_{x_1}=U(x_2-x_1)\Phi_{x_1}
\end{align}
without the level truncation. Noting
\begin{align}
U(x)(\alpha_{-1}^{25})^nc_1|0\rangle&=\left(\alpha_{-1}^{25}+\frac{x}{\sqrt{2}}\right)^n
e^{-\frac{1}{4}x^2}\exp\left(-\frac{x}{\sqrt{2}}\alpha_{-1}^{25}\right)c_1|0\rangle
\nonumber\\
&=e^{-\frac{1}{4}x^2}\left[
\left(\frac{x}{\sqrt{2}}\right)^nc_1|0\rangle+\left(
\frac{n}{\sqrt{2}}x-\left(\frac{x}{\sqrt{2}}\right)^{n+1}
\right)\alpha_{-1}^{25}c_1|0\rangle+\cdots
\right],
\end{align}
and using the expansion of the tachyon vacuum in the original theory ($x=0$):
\begin{align}
\Phi_0=\sum_{m\ge 0}a_s^{(m)}(\alpha_{-1}^{25})^{2m}c_1|0\rangle+\cdots,
\label{Phi0=}
\end{align}
where we have used the fact that $\Phi_0$ is twist even, we have
\begin{align}
\Phi_x&=U(x)\Phi_0
\nonumber\\
&=e^{-\frac{1}{4}x^2}\left[
\sum_{m\ge 0}a_s^{(m)}\left(\frac{x}{\sqrt{2}}\right)^{2m}c_1|0\rangle
+
\sum_{m\ge 0}a_s^{(m)}\left(
\sqrt{2}mx-\left(\frac{x}{\sqrt{2}}\right)^{2m+1}
\right)\alpha_{-1}^{25}c_1|0\rangle+\cdots
\right].
\label{Phix=UxPhi0}
\end{align}
Hence, if we do not truncate the level, 
the $x$-dependence of the tachyon field $t_0$ and the massless field $a_s$
is given by
\begin{align}
t_0&=e^{-\frac{1}{4}x^2}\sum_{m\ge 0}a_s^{(m)}\left(\frac{x}{\sqrt{2}}\right)^{2m},
\\
a_s&=e^{-\frac{1}{4}x^2}\sum_{m\ge 0}a_s^{(m)}\left(
\sqrt{2}mx-\left(\frac{x}{\sqrt{2}}\right)^{2m+1}
\right).
\label{as_infty}
\end{align}
It is necessary to know all coefficients $a_s^{(m)}$ ($m=0,1,2,\cdots$) in (\ref{Phi0=})
in order to obtain the exact form of (\ref{as_infty}).
However, this is impossible because 
no explicit expression of the exact solution in the Siegel gauge 
is yet known.
Instead, let us use the level-truncated numerical solution
in the original theory to obtain an approximate expression for (\ref{as_infty}).
Such a function $a_s$ with  (\ref{as_infty}) can be compared to
the plot using numerical data as given in Figs.~\ref{fig:L4as_x},
\ref{fig:L6as_x}
and \ref{fig:L8as_x}. (The left plot in each figure is from Fig.~\ref{as_x_comp}.)
With increasing truncation level, the two plots get closer. 
These plots seem to imply that level truncation can be a good approximation 
to obtain $a_s$ as a function of $x$ using numerical data.
If we use numerical configurations of the tachyon vacuum solution in the
Siegel gauge in the original theory ($x=0$), we obtain 
$a_s$ of the form (\ref{as_infty})
as a function of $x$ at each truncated level.
Using these functions, 
we can see
the maximum of $a_s$, $\max(a_s)$, and 
the value of $x$, $x_{\rm cr}$, which give $\max(a_s)$,
as in Figs.~\ref{fig:maxas_L} and \ref{fig:xcr_L}, respectively.
With the numerical data up to $L=26$ obtained in \cite{Kishimoto:2011zza},
we have extrapolated values for $L=\infty$:
$\max(a_s)=0.3118$ and $x_{\rm cr}=-1.4986$,
using a fitting function of the form $\sum_{k=0}^{13}c_k/L^k$.
We note that both values are finite.

\section{On the gauge invariant overlaps of numerical solutions in the $M$-branch
\label{sec:GIO}
}

\begin{figure}[H]
\begin{center}
\includegraphics[width=9.5cm]{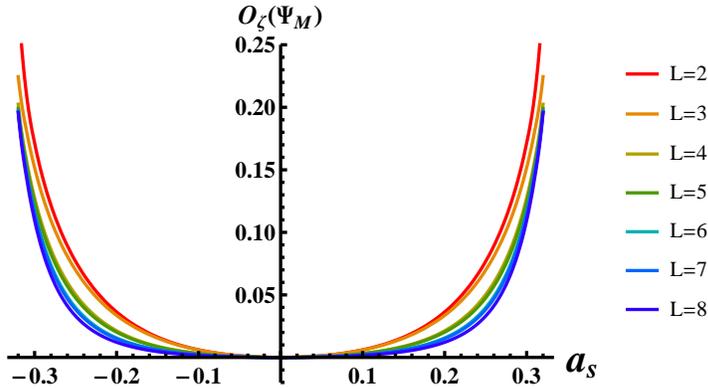}
\caption{
The gauge invariant overlap with the graviton for the $M$-branch in the Siegel gauge 
in the truncation level $L=2,3,\cdots,8$.
\label{fig:ozeta_m}}
\end{center}
\end{figure}

\begin{figure}[h]
\begin{center}
\includegraphics[width=9.5cm]{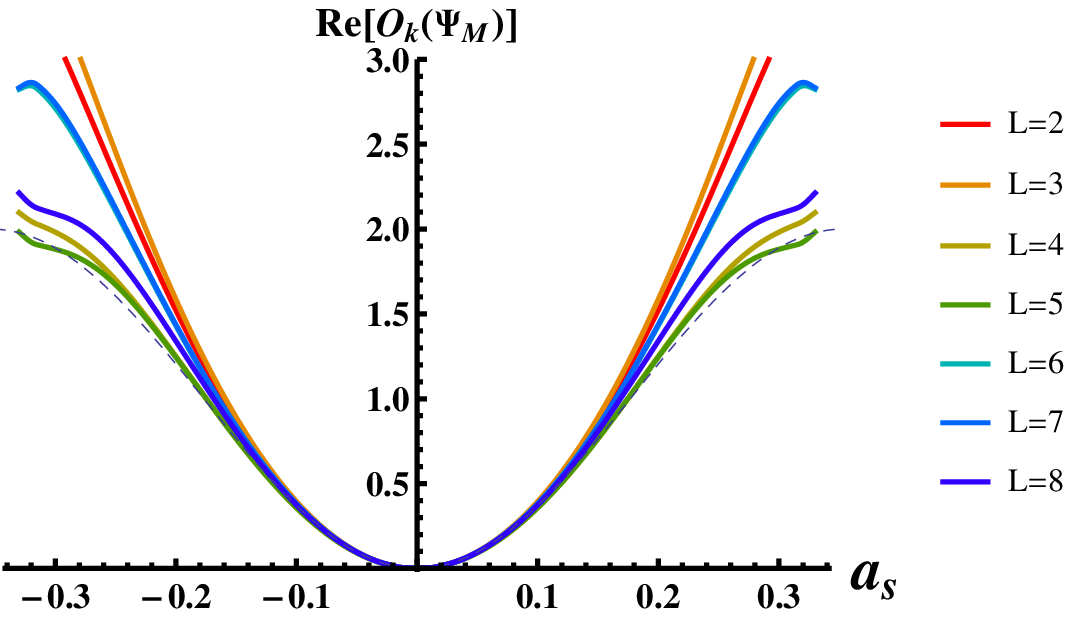}
\caption{
The real part of the gauge invariant overlap with the closed tachyon for
 the $M$-branch in the Siegel gauge 
in the truncation level $L=2,3,\cdots,8$.
The dashed line shows $1-\cos(\frac{4\pi}{\sqrt{2}}a_s)$.
\label{fig:okr_m}}
\end{center}
\end{figure}

\begin{figure}[h]
\begin{center}
\includegraphics[width=10cm]{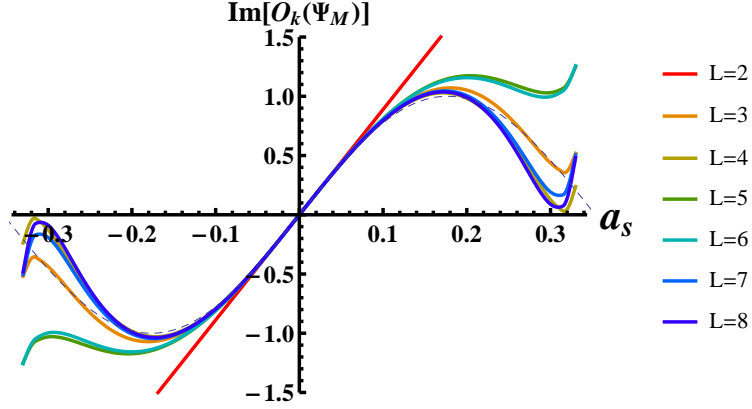}
\caption{
The imaginary part of the gauge invariant overlap with the closed
 tachyon for the $M$-branch in the Siegel gauge 
in the truncation level $L=2,3,\cdots,8$.
The dashed line shows $\sin(\frac{4\pi}{\sqrt{2}}a_s)$.
\label{fig:oki_m}}
\end{center}
\end{figure}

In the original $Q_{\rm B}$ theory, the gauge invariant overlaps for 
the configurations of the $M$-branch $\Psi_M$ in the Siegel gauge, which
correspond to the upper branch for each truncation level in
Fig.~\ref{fig:szL7}, are evaluated as Figs.~\ref{fig:ozeta_m},
\ref{fig:okr_m} and \ref{fig:oki_m}.
{}From Fig.~\ref{fig:ozeta_m}, we observe that $O_{\zeta}(\Psi_M)$
approaches $0$ for a fixed value of $a_s$ with increasing level.
This result is consistent with Fig.~\ref{fig:szL7} and the relation with the vacuum energy \cite{Baba:2012cs}.
{}From Figs.~\ref{fig:okr_m} and \ref{fig:oki_m},  
${\rm Re}(O_{\zeta}(\Psi_M))$ and  ${\rm Im}(O_{\zeta}(\Psi_M))$
nontrivially depend on $a_s$
and {\it roughly} approach 
$1-\cos(ca_s)$ and $\sin(ca_s)$,
respectively with increasing level, where $c$ is 
an appropriate constant.

Here, we would like to speculate on the possibility of a relationship between the above 
numerical result and the analytic result given in \cite{KT}.
The identity-based marginal solution $\Psi_0$(\ref{Psi0id}) is expected to
correspond to the numerical solution of the $M$-branch in the Siegel gauge.
As for the parameter, the value of the massless field $a_s$ corresponds to $f$
given in (\ref{f=intF}).
Let us consider the relationship to the parameter $x$,
which is proportional to $f$, i.e. $f=(-2/\pi)x$, more explicitly.
The massless field, or the coefficient of $\alpha_{-1}^{25}c_1|0\rangle$,
 is included only in the first term of $\Psi_0$(\ref{Psi0id}).
We can expand it as follows:
\begin{align}
&-\int_{C_{\rm left}}\frac{dz}{2\pi i}\frac{i}{2\sqrt{\alpha'}}
F(z)c(z)\partial X^{25}(z)I
=\frac{x}{\sqrt{2}}\int_{C_{\rm left}}\frac{dz}{2\pi i}\frac{z+z^{-1}}{z}\sum_{n,m}c_n\alpha^{25}_mz^{-n-m}|I\rangle
\nonumber\\
&=\frac{x}{\sqrt{2}}
\left(
\frac{8}{3\pi}(c_1+c_{-1})+
\sum_{k=1}^{\infty}\frac{2}{\pi}\left(\frac{1}{1-4k^2}-\frac{1}{1-4(k+1)^2}\right)
(c_1-c_{-1})
\right)\alpha^{25}_{-1}|I\rangle+\cdots
\nonumber\\
&=\frac{\sqrt{2}}{\pi}x\,\alpha^{25}_{-1}c_1|0\rangle+\cdots,
\end{align}
where use has been made of the relations
$\alpha^{\mu}_n|I\rangle=-(-1)^n\alpha^{\mu}_{-n}|I\rangle$
and
$(c_{2k+1}-c_{-2k-1})|I\rangle=(-1)^k(c_1-c_{-1})|I\rangle$.
Therefore, we expect the correspondence of
the parameters between the solutions to the original theory to be
 $a_s\sim (\sqrt{2}/\pi)x$.

In this context, we may expect that the numerical solution
with the parameter $a_s$, which we denote as $\Psi_M(a_s)$,
is gauge equivalent to the identity-based marginal solution $\Psi_0(x)$
(\ref{Psi0id}) 
with the parameter $x\simeq (\pi/\sqrt{2})a_s$.
If this expectation is valid, the gauge invariant overlap for them
should be
\begin{align}
&O_V(\Psi_M(a_s))\simeq O_V(\Psi_0(x=(\pi/\sqrt{2})a_s)).
\label{O_V=O_V}
\end{align}
On the right-hand side of the above equation, from the result in \cite{KT}, we have
$O_{\zeta}(\Psi_0(x))=0$ for the  graviton
and $O_k(\Psi_0(x))=1-e^{-4ix}$ for the closed tachyon,
which are roughly consistent with 
the numerical results in Figs.~\ref{fig:ozeta_m},
\ref{fig:okr_m} and \ref{fig:oki_m}.
Here, we should note that the parameter $x$ in $\Psi_0$ does not have
a finite bound and we can take any large value of $|x|$ from the viewpoint of the solution to the equation of motion.
However, the parameter $a_s$ seems to have a finite bound and 
we can take $a_s$ only in $|a_s|\lesssim 0.3$ as is seen from Fig.~\ref{fig:szL7}.
In this sense, the assumption $x\simeq(\pi/\sqrt{2})a_s$ cannot 
be justified for large values of $|a_s|$ such as $|a_s|> 0.3$.
Therefore, we do not have a definite conclusion on the gauge equivalence
between the $M$-branch numerical solution and 
the identity-based marginal solution $\Psi_0$.

\section{Concluding remarks
\label{sec:RM}
}

We have constructed numerical solutions using the conventional level
truncation method in the theory of $Q'$
obtained by expanding the string field 
around an identity-based marginal solution, which has one parameter, $x$:\\
\noindent
(a) We have constructed tachyon vacuum solutions in the Siegel gauge and
the Landau gauge.
With increasing level, the values of the action at the solutions 
approach a D-brane tension in the wider range of the parameter $x$.
This suggests that the energy of the identity-based marginal solution
vanishes as in the case of the solution with $K'Bc$ algebra
\cite{Inatomi:2012nv}, which satisfies the other gauge condition.\\
\noindent
(b) We have constructed the $M$-branch and the $V$-branch in the theories of
various values of $x$ in the Siegel gauge.
The values of the potential approach zero for higher levels 
for small values of the massless field $|a_s|$ 
in the $M$-branch and 
the potential form roughly moves in the horizontal direction according
to the values of $x$.
However,
it turns out that there seems to exist a finite bound for the value
of the massless field $a_s$ in the theory of $Q'$ with
any value of $x$
as in the original theory of $Q_{\rm B}$ ($x=0$),
which was observed in previous work \cite{Sen:2000hx}.\\
\noindent 
(c) We have evaluated the gauge invariant overlaps with
the graviton and the closed tachyon for the constructed numerical
solutions.
For the numerical tachyon vacuum solution, 
they approach the same $x$-dependence 
as the analytic ones \cite{Inatomi:2012nv}
with increasing truncation level.

For the numerical tachyon vacuum solutions $\Phi_x$ in the Siegel gauge,
we have checked the remaining part of the equation of motion
(or the BRST invariance of the gauge fixed solutions)
for consistency in appendix \ref{sec:BRST}.
As the truncation level is increased, the vacuum energy $E$
of $\Phi_x$, which is normalized by a D-brane tension, seems to become $-1$
for {\it any} value of $x$, as in Fig.~\ref{fig:Sx}.
Actually, this can be justified as follows.
Without level truncation, 
the solution in the Siegel $\Phi_x$ can be 
related to the tachyon vacuum solution in the Siegel $\Phi_0$
in the original $Q_{\rm B}$ theory $(x=0)$ as $\Phi_x=U(x)\Phi_0$
(\ref{Phix=UxPhi0}).
Noting the relations involving $U(x)$,
 (\ref{L'def}) and (\ref{UxinA*B}),
we find that
the value of the action (\ref{S'Phi}) 
does not depend on $x$, namely, $S'[\Phi_x]=S[\Phi_0]$.
Additionally, it is well confirmed that
the normalized vacuum energy of $\Phi_0$ should be $-1$
using the level truncation method.
In a similar manner, we can justify
our numerical evidence of $O_{\zeta}(\Phi_x)=1$ 
and $O_k(\Phi_x)=e^{-4ix}$
in \S \ref{sec:EGIO}:
Without the level truncation, we can show that
$O_{\zeta}(U(x)\Phi_0)=O_{\zeta}(\Phi_0)$
and 
$O_k(U(x)\Phi_0)=e^{-4ix}O_k(\Phi_0)$ using explicit expression of the 
gauge invariant overlap. 
Then, we have the equality $O_{\zeta}(\Phi_0)=O_k(\Phi_0)$
\cite{Kawano:2008ry} and its value should be $1$, as was numerically
checked in \cite{Kishimoto:2011zza}.

\section*{Acknowledgements}
The work of I. K. and T. T. is supported by a JSPS Grant-in-Aid for
Scientific Research (B) (\#24340051). 
The work of I. K. is supported in part by
a JSPS Grant-in-Aid for Young Scientists (B) (\#25800134).
The work of I. K. was supported
partly by a Grant for Promotion of Niigata University Research Projects
and partly by a Grant-in-Aid for Research Project from Institute of
Humanities, Social Sciences and Education, Niigata University.
The numerical computation in this work was partly
carried out at the Yukawa Institute Computer Facility.
\appendix

\section{On the BRST invariance of the numerical solutions
\label{sec:BRST}}

\begin{figure}[h]
\parbox{.471\textwidth}{
\begin{center}
\includegraphics[width = 8cm]{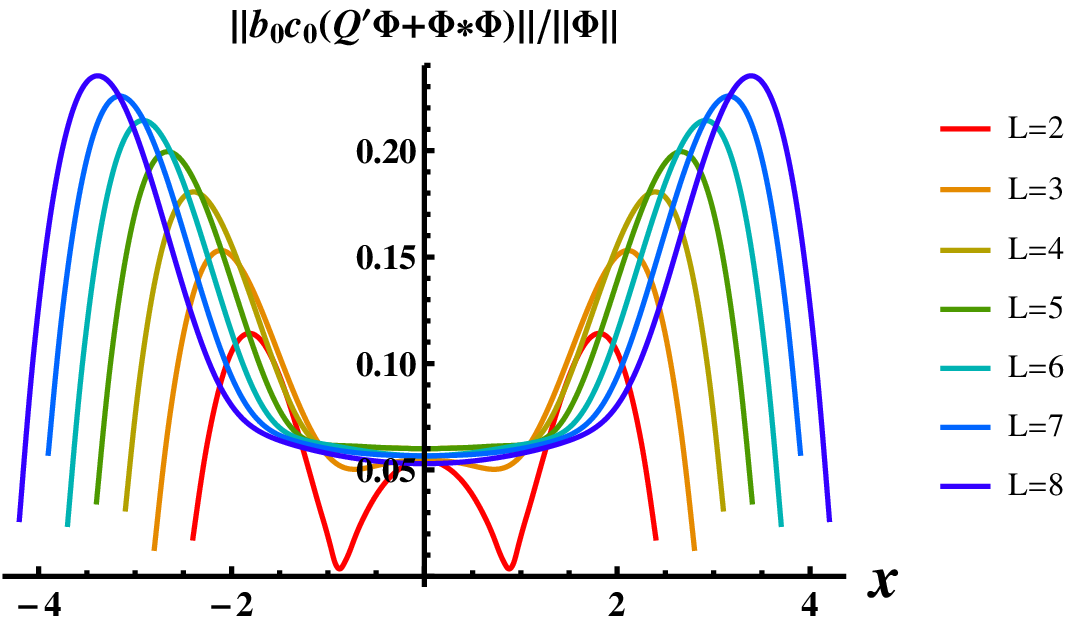}
\caption{
Plots of (\ref{BRS_norm}) in the level  $L=2,3,\cdots, 8$ truncation. 
\label{fig:brs_normL2toL8}
}
\end{center}
}
\hfill
\parbox{.471\textwidth}{
\begin{center}
\includegraphics[width = 8cm]{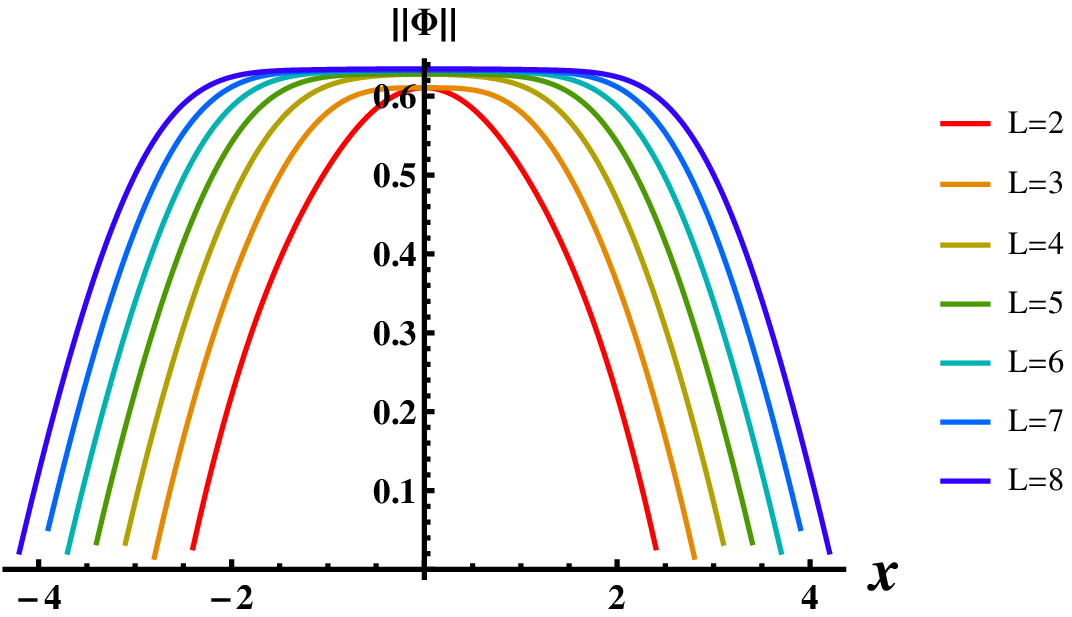}
\caption{
Plots of $\|\Phi\|$ in Fig.~\ref{fig:brs_normL2toL8}
in the level $L=2,3,\cdots, 8$ truncation. 
\label{fig:normL2toL8}
}
\end{center}
}
\end{figure}

We have constructed numerical solutions to eqs.~(\ref{G-cond})
and (\ref{P2EOM}).
However, initially, we would like to construct 
the solution to the equation of motion
 $Q'\Phi+\Phi*\Phi=0$.
Therefore, as a consistency check,  we evaluate the remaining part of the
equation of motion, namely, 
\begin{align}
{\rm bpz}({\cal P}_1)(Q'\Phi+\Phi*\Phi)=0,
\end{align}
which corresponds to the BRST invariance of the gauge fixed solution \cite{Hata:2000bj}.
In the case of the Siegel gauge, we have ${\rm bpz}({\cal P}_1)=b_0c_0$
and we evaluate
\begin{align}
 \frac{\|b_0c_0(Q'\Phi+\Phi*\Phi)\|}{\|\Phi\|},
\label{BRS_norm}
\end{align}
numerically.
In the above, the norms of string fields with ghost number 1 in the
denominator and ghost number 2 in the numerator
are defined for an orthonormalized basis
with respect to the BPZ inner product as in \cite{Kishimoto:2009cz}.
For the numerical solutions to eqs.~(\ref{G-cond}) and (\ref{P2EOM})
in the Siegel gauge, which correspond to Fig.~\ref{fig:Sx},
the ratio of norms  (\ref{BRS_norm})
is evaluated as Fig.~\ref{fig:brs_normL2toL8}.
\noindent 
Roughly speaking, the value of (\ref{BRS_norm}) remains ``small'' with
 increasing truncation level.
Although there are peaks for large $|x|$ in each plot in
Fig.~\ref{fig:brs_normL2toL8},
they are related to the decline of 
the denominator of (\ref{BRS_norm}), i.e., 
the norm of the configuration $\|\Phi\|$, as in
Fig.~\ref{fig:normL2toL8}.
Next, let us see the coefficient of $c_{-2}c_1|0\rangle$ in
$b_0c_0(Q'\Phi+\Phi*\Phi)$
as one of the component fields.
\begin{figure}[H]
\parbox{.471\textwidth}{
\begin{center}
\includegraphics[width = 8cm]{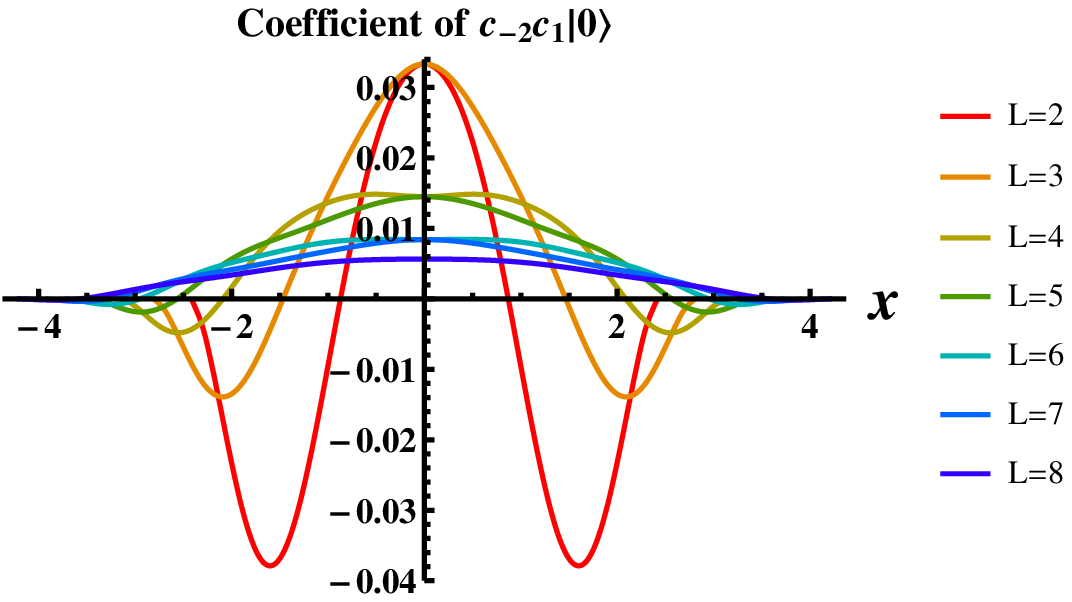}
\caption{
Plots of the coefficient of $c_{-2}c_1|0\rangle$
in $b_0c_0(Q'\Phi+\Phi*\Phi)$
in the level $L=2,3,\cdots, 8$ truncation. 
\label{fig:brscm2SxL2toL8}
}
\end{center}
}
\hfill
\parbox{.471\textwidth}{
\begin{center}
\includegraphics[width = 8cm]{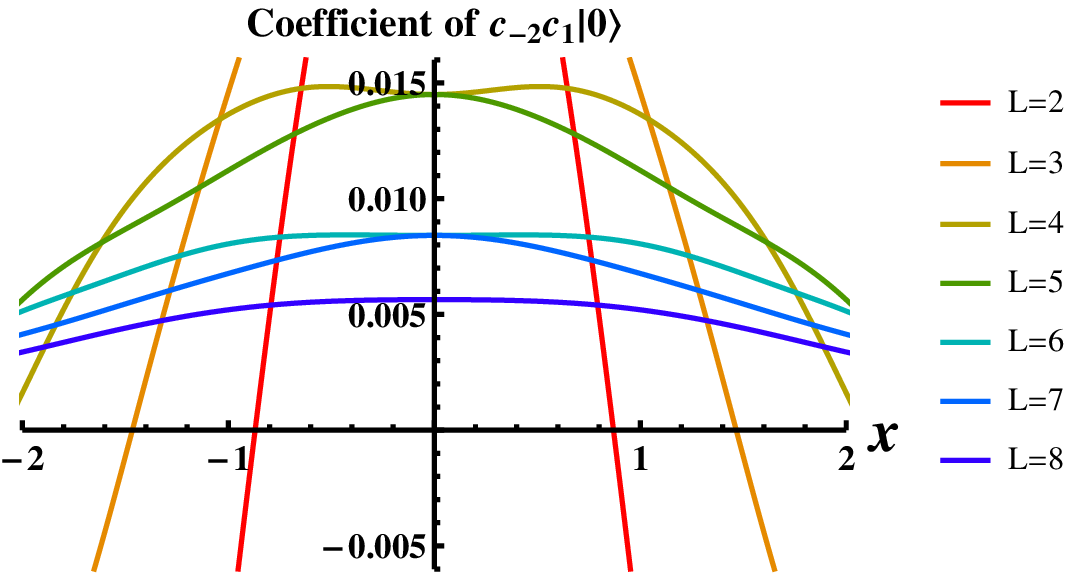}
\caption{
Enlargement of Fig.~\ref{fig:brscm2SxL2toL8}
\label{fig:brscm2SxL2toL8s}
}
\end{center}
}
\end{figure}
\noindent From Figs.~\ref{fig:brscm2SxL2toL8} and \ref{fig:brscm2SxL2toL8s},
the coefficient of $c_{-2}c_1|0\rangle$ 
gets closer to zero as the truncation level increases
for each value of $x$.
Therefore, we expect that all coefficients of $b_0c_0(Q'\Phi+\Phi*\Phi)$
approach zero in infinite level limit,
although the norm convergence might be slow.

The above results imply that our numerical solutions to eqs.~(\ref{G-cond})
and (\ref{P2EOM}), which correspond to Fig.~\ref{fig:Sx},
can be consistently regarded as approximate solutions to 
the equation of motion $Q'\Phi+\Phi*\Phi=0$.


\end{document}